\documentclass[aps,prx,superscriptaddress,twocolumn,longbibliography]{revtex4-2}
\usepackage{graphicx, color, graphpap}
\usepackage{enumitem}
\usepackage{amssymb}
\usepackage{amsthm}
\usepackage{amsmath}
\usepackage{dsfont}
\usepackage{mathtools}
\usepackage{soul}
\usepackage{xcolor}
\usepackage{colortbl}
\usepackage{booktabs}
\usepackage{multirow}
\usepackage[colorlinks=true,citecolor=blue,linkcolor=magenta]{hyperref}
\usepackage[T1]{fontenc}
\usepackage{float}

\usepackage{thmtools,thm-restate}
\usepackage{verbatim}
\usepackage{titlesec}

\usepackage{graphicx}

\usepackage{bbm}

\usepackage[lined,boxed,ruled]{algorithm2e}

\newcommand{\bra}[1]{\langle#1|}

\newcommand{\ket}[1]{|#1\rangle}

\definecolor{skyblue}{RGB}{135,206,235}

\newcommand{\resetAppendixCounters}[1]{%
  \setcounter{lemma}{0}
  \renewcommand{\thelemma}{#1.\arabic{lemma}}
  \setcounter{proposition}{0}
  \renewcommand{\theproposition}{#1.\arabic{proposition}}
  \setcounter{theorem}{0}
  \renewcommand{\thetheorem}{#1.\arabic{theorem}}
  \setcounter{corollary}{0}
  \renewcommand{\thecorollary}{#1.\arabic{corollary}}
  \setcounter{definition}{0}
  \renewcommand{\thedefinition}{#1.\arabic{definition}}
  \setcounter{subsection}{0}
  \renewcommand{\thesubsection}{#1.\arabic{subsection}}

  \renewcommand{\theequation}{#1.\arabic{equation}}
  \renewcommand{\thefigure}{#1.\arabic{figure}}
  \renewcommand{\theHfigure}{#1.\arabic{figure}} 
  \setcounter{figure}{0}
  \setcounter{equation}{0}
  
  \setcounter{algocf}{0}
  \renewcommand{\thealgocf}{#1.\arabic{algocf}}
}

\begin{document}

\title{Towards Ultimate Accuracy in Quantum Multi-Class Classification: A Trace-Distance Binary Tree AdaBoost Classifier}

\author{Xin Wang}

\affiliation{
	Department of Automation, Tsinghua University, Beijing, 100084, P. R. China
}

\author{Yabo Wang}
\email{wangyabo@swjtu.edu.cn}
\affiliation{
	School of Mathematics, Southwest Jiaotong University, Chengdu, 611756, China
}

\author{Rebing Wu}
\email{rbwu@tsinghua.edu.cn}

\affiliation{
	Department of Automation, Tsinghua University, Beijing, 100084, P. R. China
}

\begin{abstract}
    We propose a Trace-distance binary Tree AdaBoost (TTA) multi-class quantum classifier, a practical pipeline for quantum multi-class classification that combines quantum-aware reductions with ensemble learning to improve trainability and resource efficiency. TTA builds a hierarchical binary tree by choosing, at each internal node, the bipartition that maximizes the trace distance between average quantum states; each node trains a binary AdaBoost ensemble of shallow variational quantum base learners. By confining intrinsically hard, small trace distance distinctions to small node-specific datasets and combining weak shallow learners via AdaBoost, TTA distributes capacity across many small submodels rather than one deep circuit, mitigating barren-plateau and optimization failures without sacrificing generalization. Empirically TTA achieves top test accuracy ($\approx $100\%) among quantum and classical baselines, is robust to common quantum errors, and realizes aggregate systems with 10000 cumulative layers and 0.2M parameters, implemented as many shallow circuits. Our results are empirical and implementable on near-term platforms, providing a resource-efficient route to scalable multi-class quantum machine learning.
\end{abstract}

\maketitle

\section{Introduction}
Quantum machine learning (QML)~\cite{biamonte2017quantum} has emerged as a promising paradigm that could potentially surpass classical approaches by leveraging quantum properties such as superposition and entanglement~\cite{zheng2023efficient} to provide representational~\cite{huang2022quantum} or generalization advantages~\cite{caro2021encoding} in machine learning tasks such as classification. In recent years, both theoretical analyses and experiments have suggested regimes where QML models can generalize well from limited data~\cite{caro2022generalization,wang2025tight}, motivating their study for practical tasks such as image classification.

However, despite their strong generalization capabilities, the trainability of parameterized quantum circuits (PQCs) in QML models severely limits their practical utility. Specifically, shallow PQCs suffer from limited expressivity and can exhibit exponentially many poor local minima~\cite{you2021exponentially,anschuetz2022quantum}, making convergence to globally optimal solutions unlikely. While overparameterization (increasing circuit depth) has been proposed to mitigate this issue~\cite{larocca2023theory}, excessively deep circuits inevitably encounter the barren plateau phenomenon, wherein both the loss function and its gradients with respect to the circuit parameters concentrate exponentially~\cite{cerezo2021cost,ragone2024lie}, rendering large-scale QML models effectively untrainable. Compounding this challenge, noise on near-term quantum devices accumulates with circuit depth, further degrading computational fidelity~\cite{wang2021noise}. Consequently, PQC design appears trapped in a Goldilocks dilemma: circuits must be neither too shallow nor too deep, but confined to a narrow intermediate regime to maintain trainability. As a result, QML models seem unable to achieve improved performance by scaling up model size, in contrast to classical large language models~\cite{zhao2023survey}.

To scale QML models to larger sizes, we employ the AdaBoost method~\cite{freund1995desicion,zhu2009multi} to ensemble multiple shallow QML models into a large-scale QML model. The core idea of AdaBoost is to combine multiple weak base classifiers that perform only slightly better than random guessing on the training set into a strong classifier that can correctly predict most data in the training set, thereby achieving better training performance. First, AdaBoost only requires base classifiers to achieve performance better than random guessing on the training set, without needing to reach a global optimum. Therefore, on one hand, shallow QML models falling into local minima is no longer a critical issue (as long as these local minima are better than random guessing); on the other hand, by choosing shallow QML models, the barren plateau phenomenon is also avoided. Additionally, shallow models prevent quantum noise from being excessively amplified due to layer accumulation. 
Furthermore, AdaBoost only requires base classifiers to be slightly better than random guessing. This weak-learning condition frees QML base learners from relying on delicately crafted PQC structures, simplifying circuit design and improving robustness.
Moreover, from the perspective of Rademacher complexity~\cite{mohri2018foundations}, the strong classifier composed via AdaBoost enjoys the same superior generalization performance as the QML models that constitute the base classifiers.

Nevertheless, AdaBoost provides trainability guarantees only for binary classification, not for the more practical multi-class setting. For binary classification, the train error bound of AdaBoost's strong classifier decreases exponentially with the number of weak classifiers, while no similar guarantee exists for multi-class problems. Consequently, for real-world multi-class datasets, the improvement of multi-class AdaBoost with QML models as base classifiers remains limited~\cite{wang2024supervised}. A natural idea to extend the training guarantees of binary classification to multi-class problems is to reduce the multi-class problem into multiple binary classification subproblems. Standard strategies for reducing multi-class problems to binary classification mainly include: (i) bitwise aggregate, which encode class labels into binary form and use multiple binary classifiers, each responsible for predicting a single bit; the final class is determined by decoding the predicted bits~\cite{dietterich1994solving}; (ii) one-versus-rest (OVR), which train one classifier per class by treating that class as positive and all other classes as negative, with prediction based on the classifier with the highest confidence for a given test instance; (iii) one-versus-one (OVO), which build a classifier for every pair of classes, treating only the two classes at a time and aggregating their decisions via voting~\cite{hastie1997classification}. We provide a more detailed introduction to these three reduction methods in App.~\ref{app:reduction_methods}. However, when applied to QML, these approaches do not exploit quantum-specific information. In particular, the intrinsic distinguishability of the class quantum states (for example, the trace distance between expected states) can have a significant impact on the trainability of quantum classifiers~\cite{wang2025limitations}. The binary classification subproblems whose quantum states are nearly indistinguishable will remain hard for any quantum classifier; in such cases even binary AdaBoost may fail to  achieve 100\% training accuracy for binary classifiers, which can propagate through reductions to degrade multi-class performance.

To address these issues, we propose a novel quantum multi-class classifier: the Trace-distance binary Tree AdaBoost (TTA) Classifier. This method partitions multi-class training sets into positive and negative subsets, then recursively applies the same binary partitioning to both positive and negative subsets, constructing a dataset binary tree until each leaf node contains samples from only a single class. Unlike previous multi-class to binary reduction methods, the TTA classifier leverages the quantum-specific property that the trace distance between expected states of two classes affects the training performance of QML models. Therefore, each partition seeks the combination with maximum trace distance among all possible class combinations to facilitate QML model training. We then apply the AdaBoost method to each binary classifier to enhance trainability, achieving superior training performance for the entire multi-class task.  By leveraging the strong generalization capability of quantum machine learning models, the overall classifier maintains excellent predictive performance on multi-class tasks.  Here, the TTA classifier achieves cumulative circuit depths of up to approximately 10,000 layers and parameter scales of 0.2M in QML models, avoiding the "U-shaped curve" exhibited by single QML models on multi-class tasks~\cite{du2023problem} and demonstrating the capabilities of large-scale QML.

For an intuitive comparison, we evaluate TTA on image and synthetic benchmarks. On the MNIST dataset, we compare against: (1) single quantum multi-class classifier, (2) quantum multi-class AdaBoost classifier, (3) quantum classifier with bitwise AdaBoost, (4) classical multi-class ResNet50, and (5) classical multi-class ViT-Small. On the synthetic dataset, we further compare against the OVO and OVR methods. The final results are summarized in Table~\ref{tab:mnist_synth_results}. In the table, we highlight the best train accuracy, best test accuracy, and the minimal model parameter count when the best test accuracy is achieved. Across these metrics, the TTA classifier consistently achieves the highest prediction accuracy with the fewest parameters compared to the competing quantum and classical approaches.

\begin{table*}[htbp]
    \centering
    \caption{Performance comparison (mean $\pm$ std) of  various methods on benchmark   datasets. M denotes $1,000,000$.}
    \label{tab:mnist_synth_results}
    \begin{tabular}{llccc}
        \toprule
        Dataset & Model / Method & Training Accuracy & Test Accuracy & Number of Parameters \\
        \midrule
        \multirow{6}{*}{MNIST}
            & Single Classifier (quantum, no aggregate)           & 89.05$\pm$0.36\% & 88.73$\pm$0.43\% & 0.00048$\pm$0.00000 M \\
            & multi-class AdaBoost (quantum, aggregate)     & 96.77$\pm$0.20\% & 96.05$\pm$0.16\% & 0.24480$\pm$0.00000 M \\
            & Bitwise AdaBoost (quantum, aggregate)        & 99.46$\pm$0.04\% & 97.30$\pm$0.07\% & 0.24000$\pm$0.00000 M \\
            & ResNet50 (classical, no ensemble)       & \textbf{100.00$\pm$0.00\%} & 97.48$\pm$0.64\% & 22.0837$\pm$0.00000 M \\
            & ViT Small (classical, no ensemble)      & \textbf{100.00$\pm$0.00\%} & 98.53$\pm$0.18\% & 25.5570$\pm$0.00000 M \\
            & \textbf{TTA (quantum, ensemble, ours)}                     & \textbf{100.00$\pm$0.00\%} & \textbf{98.71$\pm$0.11\%} & \textbf{0.23433$\pm$0.01269 M} \\
        \midrule
        \multirow{3}{*}{Synthetic}
            & OVO (quantum, aggregate)                     & 75.75$\pm$2.17\% & 76.52$\pm$2.04\% & 0.03607$\pm$0.00353 M \\
            & OVR (quantum, aggregate)                     & \textbf{100.00$\pm$0.00\%} & \textbf{100.00$\pm$0.00\%} & 0.05256$\pm$0.00390 M \\
            & \textbf{TTA (quantum, ensemble, ours)}                     & \textbf{100.00$\pm$0.00\%} & \textbf{100.00$\pm$0.00\%} & \textbf{0.02405$\pm$0.00255 M} \\
        \midrule
        \bottomrule
    \end{tabular}
\end{table*}

\section{Preliminaries}\label{sec:preliminaries}

This section provides a brief overview of the key foundations for our study. In Subsec.~\ref{subsec:supervised_qml}, we introduce the supervised QML settings. In Subsec.~\ref{subsec:binary_adaboost}, we summarize the binary AdaBoost algorithm used as a building block in our pipeline.
\subsection{Supervised QML}\label{subsec:supervised_qml}
In supervised QML, we denote the input space by $\mathcal{X}$ and the label space by $\mathcal{Y}$, and let $\mathcal{D}$ be an unknown joint distribution over $\mathcal{X}$ and $\mathcal{Y}$. The input space $\mathcal{X}$ is uniformly defined as the space of quantum state density matrices: for native quantum data, the quantum state $\rho$ directly belongs to $\mathcal{X}$; for classical data $\boldsymbol{x}$, we prepare its corresponding quantum state $\rho(\boldsymbol{x}) \in \mathcal{X}$ through an encoding mapping. The training set $S = \{(\rho^{(m)},y^{(m)})\}_{m=1}^M$ contains $M$ samples, each independently sampled from $\mathcal{X} \times \mathcal{Y}$. 
The output of a single QML model is defined as the expectation value of the observable $O$ on the state evolved by a PQC $U(\boldsymbol{\theta})$: $h(\rho,\boldsymbol{\theta}) = \operatorname{Tr}\left[O U(\boldsymbol{\theta}) \rho U^{\dagger}(\boldsymbol{\theta})\right]$. The hypothesis learned on training set $S$ is $h_S(\rho) = h(\rho,\boldsymbol{\theta}^{*})$, where $\boldsymbol{\theta}^{*}$ is the parameter selected during the training process. In this paper, we choose the observable to be the Pauli $Z$ operator on the first qubit, denoted as $Z_{1}$. The architecture of the PQC $U(\boldsymbol{\theta})$ is illustrated in Fig.~\ref{fig:arch}.

During training, we typically employ a differentiable loss function $L$ as the optimization objective to guide model parameter learning. To evaluate the model's predictive performance, we measure the discrepancy between model predictions and true labels using a per-sample risk function $r(\cdot,\cdot)$, and define the empirical error of hypothesis $h_S$ on dataset $S' = \{(\rho'^{(m)},y'^{(m)})\}_{m=1}^{M}$ as 
$$
\widehat{R}_{S'}(h_S) = \frac{1}{M} \sum_{m=1}^M r(h_S(\rho'^{(m)}),y'^{(m)}).
$$
When $S' = S$, this is the train error, and $1 - \widehat{R}_S(h_S)$ is the train accuracy; when dataset $S'$ contains completely unseen data drawn from the same distribution as training set $S$, we call $\widehat{R}_{S'}(h_S)$ the test error. The prediction error of hypothesis $h_S$ learned from training set $S$ on distribution $\mathcal{D}$ is defined as:
$$
R(h_S) = \underset{(\rho,y) \sim \mathcal{D}}{\mathbb{E}}[r(h_S(\rho),y)].
$$
The prediction error cannot be directly measured; in practice, test error is generally used as a proxy for prediction error. The fundamental goal of machine learning is to ensure that models performing well on training data also perform well on unseen data, this property is called generalization capability.

\subsection{Binary AdaBoost}\label{subsec:binary_adaboost}

\begin{algorithm}[htbp]
    \caption{Binary Quantum AdaBoost}\label{algo:binary_adaboost}
    \SetKwInOut{Input}{Input}
    \SetKwInOut{Output}{Output}
    \SetKwProg{Fn}{}{:}{}
    \SetKw{KwTo}{to}
    \SetKwComment{tcp}{// }{}
    
    \Input{
        Training data $S = \left\{ \left( \rho^{(m)}, y^{(m)} \right) \right\}_{m=1}^M$,  
        where $\rho^{(m)}$ is a quantum state (density matrix), $y^{(m)} \in \{-1, +1\}$; \\
        Max boosting rounds $T \in \mathbb{N}$; \\
        Early stopping tolerance $\tau = 0.005$.
    }
    \Output{Final ensemble AdaBoost binary classifier $H_S(\cdot)$ }
    
    \BlankLine
    Initialize uniform weights: $\boldsymbol{w}_1(m) \gets \frac{1}{M}$ for $m=1,\ldots,M$\\
    
    \For{$t \gets 1$ \KwTo $T$}{

        Train base classifier $h_S^{(t)}: \rho \mapsto \{-1, +1\}$ on $S$ w.r.t. $\boldsymbol{w}_t$\\
        Compute weighted error: $
            \epsilon_t \gets \sum_{m=1}^M \boldsymbol{w}_t(m) \cdot \mathbb{I}\big( y^{(m)} \neq h_S^{(t)}(\rho^{(m)}) \big)
        $
        \tcp{Terminate if not better than random}
        \If{$\epsilon_t \geq 1/2$}{
            break \tcp*{End boosting}
        }
        Compute bound: $
            \gamma_t \gets \exp\bigg( -2 \sum_{i=1}^{t} \left( \frac{1}{2} - \epsilon_i \right)^2 \bigg)
            $ \\
        \If{$\gamma_t < \tau$}{
            break \tcp*{Enough confidence}
        }
        Update coefficient: $\alpha_t \gets \dfrac{1}{2} \log \dfrac{1-\epsilon_t}{\epsilon_t}$
    
        Compute normalization factor: $Z_t \gets 2 \sqrt{\epsilon_t (1-\epsilon_t)}$
    
        \For{$m \gets 1$ \KwTo $M$}{
            Update weights: $\boldsymbol{w}_{t+1}(m) \gets [\boldsymbol{w}_t(m) \exp\big(-\alpha_t\, y^{(m)} h_S^{(t)}(\rho^{(m)})\big)] / Z_t$
        }
    }

    $H_S(\rho) = \operatorname{sgn}\left( \sum_{t=1}^{T} \frac{\alpha_t}{\|\boldsymbol{\alpha}\|_{1}}\, h_S^{(t)}(\rho) \right)$, where $\|\boldsymbol{\alpha}\|_{1} = \sum_{t=1}^{T} \alpha_t$ \\
    \Return $H_S$
\end{algorithm}

Binary AdaBoost (shown in Algo.~\ref{algo:binary_adaboost}) is an adaptive boosting algorithm that improves training performance by iteratively training multiple weak classifiers and combining them into a strong classifier, with theoretical train error guarantees. 
When training the $t$-th base classifier on training set $S = \{(\rho^{(m)},y^{(m)})\}_{m=1}^{M}$, the weight of sample $(\rho^{(m)},{y}^{(m)})$ is $\boldsymbol{w}_t(m)$. We employ the hinge loss~\cite{rosasco2004loss} to make the  $t$-th base classifier's prediction results closer to the labels:
\begin{equation}
    \label{eq:hinge_loss}
\begin{aligned}
    \mathcal{L}_t(\boldsymbol{\theta} ; S)= \sum_{m=1}^{M} \boldsymbol{w}_t(m)  \max \left\{0,1-y^{(m)} h(\rho^{(m)},\boldsymbol{\theta})\right\} .
\end{aligned}
\end{equation}
After training, we obtain the $t$-th base classifier $h_S^{(t)}(\rho) = h(\rho,\boldsymbol{\theta}^{*})$, where $\boldsymbol{\theta}^{*}$ is the parameter selected during the training process. We record the error of $h_S^{(t)}$ as $\epsilon_{t} = \sum_{m=1}^{M}\boldsymbol{w}_t(m) \mathbbm{1} \left( y^{(m)} \neq h_S^{(t)}\left( \rho^{(m)} \right) \right) $, where $\mathbbm{1}$ is the indicator function that equals 1 if the condition is true and 0 otherwise. With the first $t$ base classifiers, we obtain the current upper bound of train error on training set $S$:
$
\gamma_t = \exp \left[ - 2 \sum_{i=1}^{t} \left( 1 / 2 - \epsilon_i \right)^2  \right].
$
As can be seen, as long as the error $\epsilon_t < 1/2$ of the $t$-th base classifier, the overall error upper bound will decrease, and the larger the gap between $\epsilon_t$ and $1/2$, the faster the overall error upper bound decreases. Moreover, the error upper bound decreases exponentially as the number of AdaBoost training rounds $t$ increases. 

After obtaining the error $\epsilon_t$, we can derive the coefficient $\alpha_t$ of the current classifier and update the weights of the training samples to $\boldsymbol{w}_{t+1}$, increasing the weights of samples that were incorrectly classified in the $t$-th round and decreasing the weights of correctly classified samples, so that the $(t+1)$-th base classifier focuses on higher-weighted samples misclassified in previous rounds. The final ensemble AdaBoost classifier is defined as $H_S(\rho) = \operatorname{sgn} ( \sum_{t=1}^T \alpha_t h_S^{(t)}(\rho) / \|\boldsymbol{\alpha}\|_{1})$, where $T$ is the total number of AdaBoost rounds, $\boldsymbol{\alpha} = [\alpha_1,\cdots,\alpha_T]^{\top}$ is the vector composed of coefficient of the $T$ base classifiers. The reason for using $\|\boldsymbol{\alpha}\|_1$ for normalization is that this ensures the term inside $\operatorname{sgn}\left(  \right) $ has a value range of $[-1,1]$ and is a convex combination of $h_S^{(t)}$. This guarantees that the entire AdaBoost method also possesses the same generalization capability as individual QML models~\cite{mohri2018foundations}.

During AdaBoost training, there are three stopping conditions: (1)When the train error upper bound $\gamma_t$ drops below $0.005$ after the $t$-th base classifier is trained, the preset goal is met and thus training terminates. (2) When the number of training rounds reaches the preset maximum number of rounds, training exits; (3) When the error of the $t$-th base classifier $\epsilon_t > 1/2$, it cannot meet the basic requirement of a weak base classifier and exits. We refer to condition (1) as a convergence situation, and conditions (2) and (3) as non-convergence situations.

\begin{figure*}
    \centering
    \includegraphics[width=0.9\textwidth]{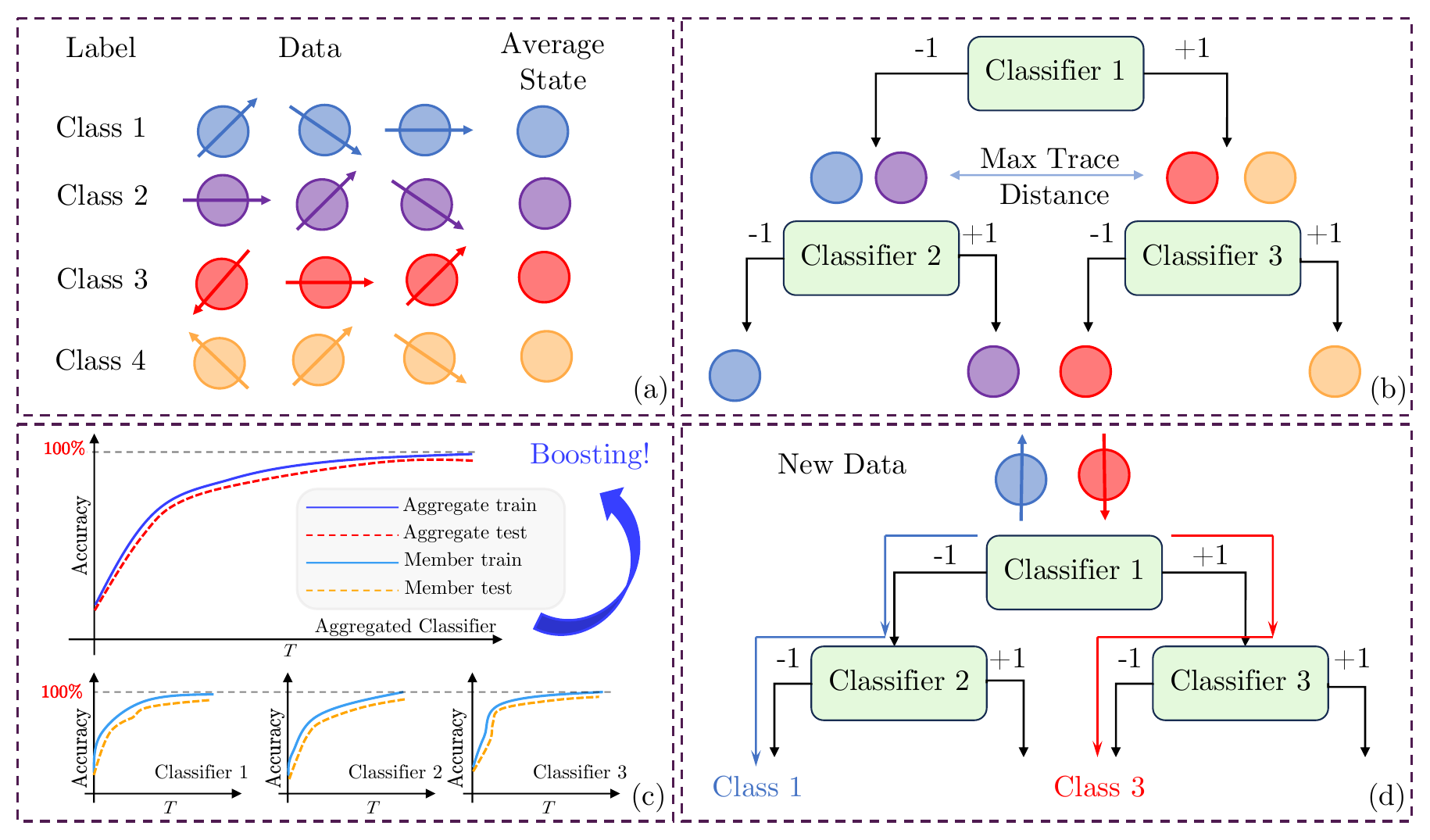}
    \caption{A schematic illustration of a trace-distance binary tree AdaBoost (TTA) multi-class classifier (a) Quantum state data of different classes and their average states. (b) Construction of a binary dataset tree by choosing splits that maximize the trace distance between the average states of two possible classes; each internal node holds a binary member classifier. (c) AdaBoost improves each member's train accuracy toward 100\% and, together with QML generalization, yields near-perfect aggregate prediction. (d) Inference proceeds from the root to a leaf along node decisions to produce the final class label.}
    \label{fig:TTA}
\end{figure*}

\section{Trace Distance Binary Tree}\label{sec:trace_distance_tree}

This section presents the TTA framework. Subsec.~\ref{subsec:dataset_partitioning} introduces the dataset partitioning strategy based on trace distance maximization. Subsec.~\ref{subsec:training_inference} describes the training and inference procedures.

\subsection{Dataset Partitioning}\label{subsec:dataset_partitioning}

In quantum binary classification problems, when the trace distance between the average quantum states of the two classes is very small, QML models struggle to correctly distinguish between them~\cite{wang2025limitations}. Furthermore, another intuition is that the more data in the training set, the harder it is for a QML model to correctly classify all samples. In the extreme case, classifying just two samples is certainly easier than classifying thousands.

Motivated by these observations, we aim to achieve two objectives when constructing binary membership classifiers: (1) maximize the trace distance between the average states of the positive and negative classes as much as possible; (2) for cases in the dataset where the trace distance between the average states of the positive and negative classes is small, minimize the number of samples in both classes.

\begin{algorithm}[htbp]
    \caption{Recursive Bipartitioning Based on Trace Distance}\label{alg:recursive_bipart}
    \SetKwInOut{Input}{Input}
    \SetKwInOut{Output}{Output}
    \SetKwFunction{RecurSplit}{RecurSplit}
    \SetKwFunction{MaxBinarySplit}{MaxBinarySplit}
    \SetKwProg{Fn}{}{:}{}
    \SetKwComment{tcp}{// }{}

    \Input{Balanced dataset $S = \{(\rho^{(m)}, y^{(m)})\}_{m=1}^M$, Label set $K$ with $|K| \geqslant 2$ classes}
    \Output{List $\mathcal{P}$ of bipartitions (paired dataset splits) at all internal nodes}

    \Fn{\RecurSplit{$S_{\textnormal{cur}}, K_{\textnormal{cur}}$}}{
        \uIf{$|K_{\textnormal{cur}}| \leq 1$}{
            \Return $\varnothing$ \tcp*{No further processing}
        }
        \ElseIf{$|K_{\textnormal{cur}}| = 2$}{
            Let $k_1, k_2$ be the two classes in $K_{\textnormal{cur}}$\;
            $S_{-} \gets \{ (\rho^{(m)}, y^{(m)}) \in S_{\textnormal{cur}} : y^{(m)} = k_1 \}$\;
            $S_{+} \gets \{ (\rho^{(m)}, y^{(m)}) \in S_{\textnormal{cur}} : y^{(m)} = k_2 \}$\;
            \Return $\{(S_{-}, S_{+})\}$\;
        } 
        \Else{
            \tcp{Step 1: Compute the average quantum state for each class}
            \ForEach{class $k$ in $K_{\textnormal{cur}}$}{
                Compute $\overline{\rho}_k = 1 / M_k \sum_{m_k=1}^{M_k} \rho^{(m_k)}$\;
            }
            \tcp{Step 2: Find bipartition maximizing trace distance in label set $K_{\textnormal{cur}}$}
            $(K_{-}, K_{+}) \gets$ \MaxBinarySplit{$\{\overline{\rho}_k\}_{k \in K_{\textnormal{cur}}}$}\;
            \tcp{Step 3: Split data by partitioned classes}
            $S_{-} \gets \left\{ (\rho^{(m)}, y^{(m)}) \in S_{\textnormal{cur}} : y^{(m)} \in K_{-} \right\}$\;
            $S_{+} \gets \left\{ (\rho^{(m)}, y^{(m)}) \in S_{\textnormal{cur}} : y^{(m)} \in K_{+} \right\}$\;
            \tcp{Step 4: Recursive calls and output of this split}
            $\mathcal{P}_{-} \gets$ \RecurSplit{$S_{-}$, $K_{-}$}\;
            $\mathcal{P}_{+} \gets$ \RecurSplit{$S_{+}$, $K_{+}$}\;
            \Return $\{(S_{-}, S_{+})\} \cup \mathcal{P}_{-} \cup \mathcal{P}_{+}$\;
        }
       
    }
    \BlankLine
    \textbf{Main routine:}\\
    $\mathcal{P} \gets$ \RecurSplit{$S$, $K$}\;
    \ForEach{$(S_{-}, S_{+}) \in \mathcal{P}$}{
        $S_{-}$: $y^{(m)} \gets -1$, $S_{+}$: $y^{(m)} \gets +1$\;
    }
    \Return $\mathcal{P}$
\end{algorithm}

Based on these two principles, we construct the TTA classifier as illustrated in Fig.~\ref{fig:TTA}. For a $K$-class training set with roughly balanced class sizes, we first partition the $K$ classes into two groups of roughly equal size (i.e., $K/2$ classes each when $K$ is even, or $\lfloor K/2 \rfloor$ and $\lceil K/2 \rceil$ classes when $K$ is odd). The partition is chosen to maximize the trace distance between the average quantum states of the two resulting groups. This process is then recursively applied to each of the resulting subsets. The detailed algorithm can be found in Algo.~\ref{alg:recursive_bipart}.

Concretely, the algorithm recursively bisects a $K$-class dataset: at each step, after splitting, it recursively processes both child subsets until each subset contains only one classes or fewer, recording every split made. With this decomposition, the two classes with the smallest trace distance are pushed as far as possible toward the leaf nodes. This ensures that at nodes with large sample sizes, the trace distance between positive and negative classes is maximized, while nodes with small trace distances between positive and negative classes have as few samples as possible. By combining this with the AdaBoost method to improve the trainability of each member classifier, we can ultimately achieve 100\% train accuracy under our experimental conditions. Additionally, due to the strong generalization ability of QML models, the test accuracy is also close to 100\%.

To clarify, consider a concrete example: suppose the set of classes is $\{1,2,3,4,5\}$ and all pairwise mean-state trace distances are given. (1) Select the partition with maximal trace distance  (e.g., this is $\{1,2,3\}$ versus $\{4,5\}$), and record the split $(S_{123},S_{45})$. (2) Recursively apply the procedure to $S_{123}$ (now $K=3$); again, suppose the maximal trace distance split is $\{1,2\}$ versus $\{3\}$, so we record $(S_{12},S_{3})$. (3) Recurse on $S_{45}$ ($K=2$): since only two classes remain, record $(S_{4},S_{5})$ and do not split further. (4) Recurse on $S_{12}$ ($K=2$): again record $(S_{1},S_{2})$ and stop. (5) Recurse on $S_{3}$ ($K=1$): with only one class, no further split or output is needed. At the end, the set of partitions is  
$$
\{(S_{123},S_{45}), (S_{12},S_{3}), (S_{4},S_{5}), (S_{1},S_{2})\}
$$ for a total of $K-1=4$ splits. For every subset in the list, new binary labels are assigned: by convention, the classes indicated on the left of the index are assigned $-1$ (negative class), and those on the right are assigned $+1$ (positive class). For example, in $(S_{123},S_{45})$, classes 1, 2, and 3 are assigned $-1$, while classes 4 and 5 are assigned $+1$.

An important component of this algorithm is the \texttt{MaxBinarySplit} procedure, which identifies the bipartition with the maximum trace distance among all possible combinations. When the number of classes $K$ is small, a brute-force approach can be used by enumerating all $\binom{K}{\left\lfloor K/2 \right\rfloor}$ possible combinations to find the partition with the largest trace distance. However, as $K$ grows, the computational complexity of this approach becomes prohibitive. To address this, we present a greedy heuristic in App.~\ref{app:other_algorithms} that partitions $K$ classes in  $\mathcal{O}(K^2 \log K)$ operations.

\subsection{Training and Inference}\label{subsec:training_inference}

\begin{figure*}
    \centering
    \includegraphics[width=0.87\textwidth]{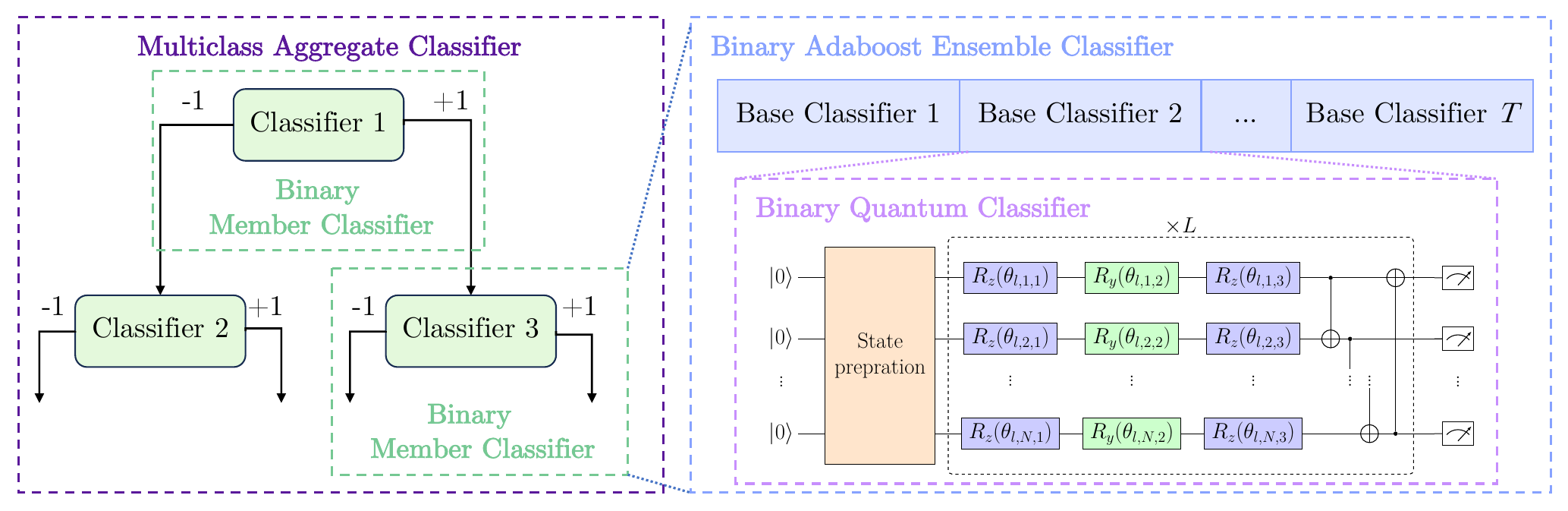}
    \caption{Schematic illustration of a multi-class aggregate classifier constructed from an ensemble of binary AdaBoost member classifiers. Each member is an AdaBoost ensemble of binary quantum base classifiers: data are encoded or prepared as quantum states, evolved by a shallow parameterized circuit (layers of single-qubit $R_z$, $R_y$, and $R_z$ rotation gates plus ring-topology CNOT gates), and measured (Pauli-Z on the first qubit) to give binary outputs.}
    \label{fig:arch}
\end{figure*}

Once the partition list $\mathcal{P}$ is obtained, we can train AdaBoost classifiers on all the binary classification subsets in $\mathcal{P}$ in parallel, ultimately forming a multi-class classifier. The partitioning of the dataset creates a binary tree structure, and correspondingly, the classifiers that classify each of these binary subsets also form a binary tree. During inference, an unlabeled sample starts at the root node of the classifier tree and traverses downward according to the outputs at each node: if the predicted label is $-1$, the sample proceeds to the left child classifier; if the predicted label is $+1$, it goes to the right child classifier. This process continues until the sample reaches a leaf node, at which point the class corresponding to that leaf node is taken as the predicted class for the sample.

As illustrated in Fig.~\ref{fig:arch}, the overall multi-classifier constructed from TTA is referred to as the aggregate classifier, while the binary classifiers at each node of the binary tree are called member classifiers. Each member classifier is essentially an ensemble classifier that combines multiple weak base classifiers using the binary AdaBoost method. Each base classifier is implemented as a quantum circuit with $N$ qubits and a depth of $L$ layers. Each layer comprises single-qubit $R_z$, $R_y$, and $R_z$ rotation gates, as well as CNOT gates arranged in a ring topology. The output is obtained by measuring the operator $Z_1$, which is the Pauli Z operator acting on the first qubit, thus forming a binary classification model. When making predictions on new data, the data is input into multiple quantum circuits; the AdaBoost method then ensembles the outputs of each quantum circuit to produce the prediction result of the member classifier. Finally, the predictions from all member classifiers are aggregated to yield the final multi-class prediction for the given input.

\section{Classification on Classical Data}\label{sec:classical}

This section evaluates the TTA classifier on classical datasets. Subsec.~\ref{subsec:mnist_dataset} introduces the MNIST dataset and experimental setup. Subsec.~\ref{subsec:performance} presents performance comparisons with various quantum and classical classifiers. Subsec.~\ref{subsec:details_tta} provides detailed analysis of the TTA classifier's behavior. Subsec.~\ref{subsec:details_compared} describes the compared methods in detail.

\subsection{MNIST dataset}\label{subsec:mnist_dataset}
\begin{figure}
    \centering
    \includegraphics[width=0.859\columnwidth]{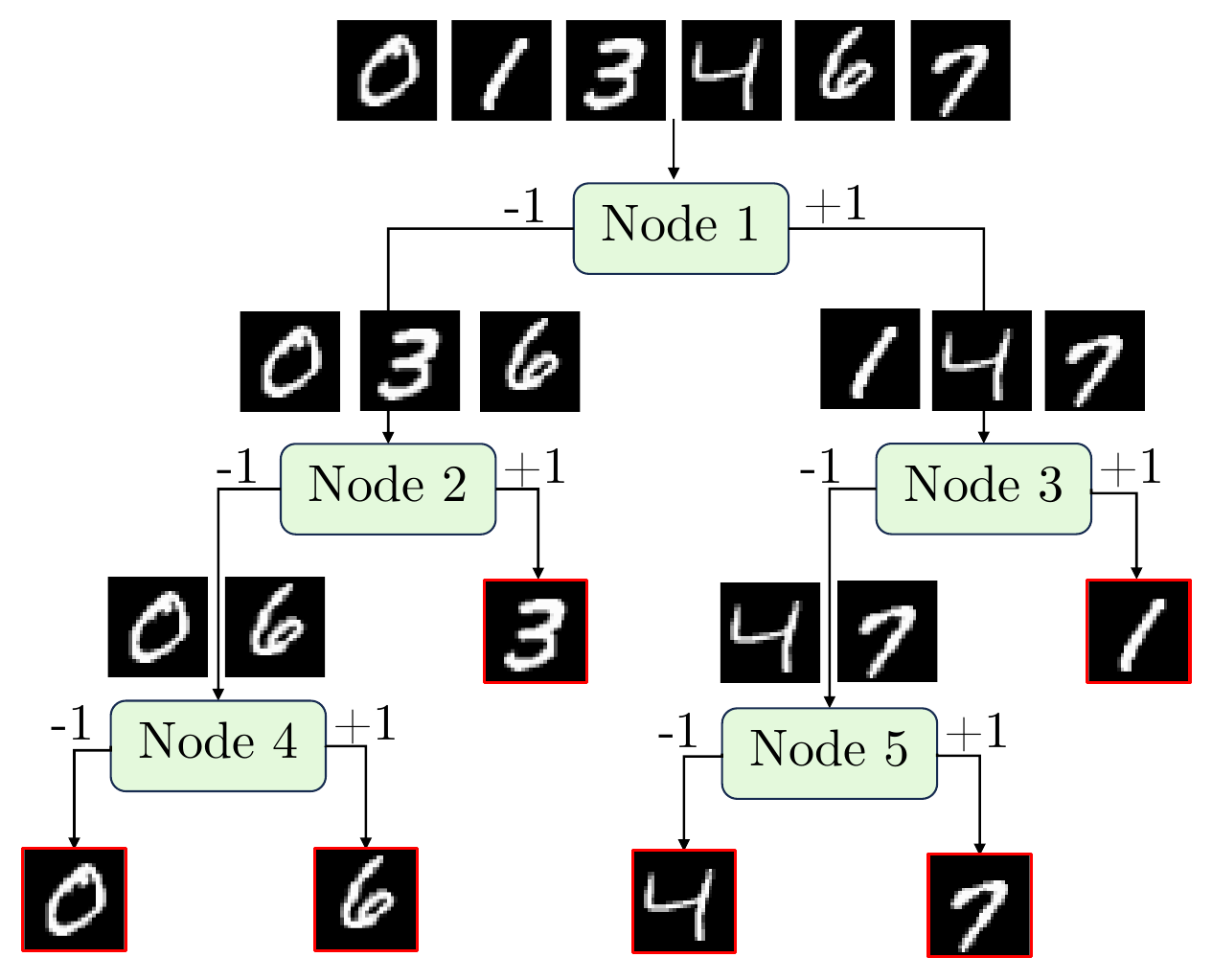}
    \caption{Trace distance tree for the MNIST dataset. Each node is a binary member classifier, and red nodes indicate leaf nodes.}
    \label{fig:mnist_tree}
\end{figure}

In this experiment, we use a subset of the MNIST dataset~\cite{lecun1998gradient} containing digits 0, 3, 6, 1, 4, and 7, forming a 6-class classification problem. We resize the MNIST images to $16 \times 16$ pixels. The training set contains 2000 samples per class, totaling 12000 training samples, while the test set contains 6093 samples. We construct the binary tree by maximizing the trace distance between the average states of negative and positive classes, with the construction results shown in Fig.~\ref{fig:mnist_tree}. Each node in the tree represents a member classifier, and the trace distance values between the negative and positive class average states corresponding to each node are listed in Table~\ref{tab:mnist_trace_distances}.

\begin{table}[h]
\centering
\caption{Trace distances between the average quantum states of the negative class (left) and positive class (right) for each node in the MNIST TTA tree.}
\label{tab:mnist_trace_distances}
\begin{tabular}{cccc}
\hline
\hline
Node & Classes  & Trace Distance & Training Samples \\
\hline
Node 1 & \{0,3,6\} vs \{1,4,7\} & 0.5672 & 12000 \\
Node 2 & \{0,6\} vs \{3\} & 0.5892 & 6000 \\
Node 3 & \{4,7\} vs \{1\} & 0.7366 & 6000 \\
Node 4 & \{0\} vs \{6\} & 0.6109 & 4000 \\
Node 5 & \{4\} vs \{7\} & 0.5593 & 4000 \\
\hline
\hline
\end{tabular}
\end{table}

To train the TTA classifier, the resized images are amplitude-encoded into an $8$-qubit quantum state and processed by the PQC illustrated in Fig.~\ref{fig:arch} with circuit depth $L=20$ layers, where each layer contains 24 trainable parameters, resulting in a total of 480 parameters per base classifier. The initial parameters of each base classifier's quantum circuit are independently sampled from a standard Gaussian distribution. We use the Pauli Z operator on the first qubit as the measurement operator and adopt hinge loss defined in Eq.~\eqref{eq:hinge_loss} as the loss function. The training is performed with a batch size of $200$ using the Adam optimizer~\cite{kingma2014adam} with a learning rate of $0.005$. We limit the maximum number of epochs to 100 and employ an early stopping strategy that halts training when the train error falls below 0.5 and does not decrease for 10 consecutive epochs, selecting the final parameters as those corresponding to the minimal train error achieved before early stopping. The effectiveness of this early stopping strategy and its ability to reduce training costs will be demonstrated in subsequent sections.

\subsection{Performance}\label{subsec:performance}
\begin{figure}[H]
    \centering
    \includegraphics[width=0.49\textwidth]{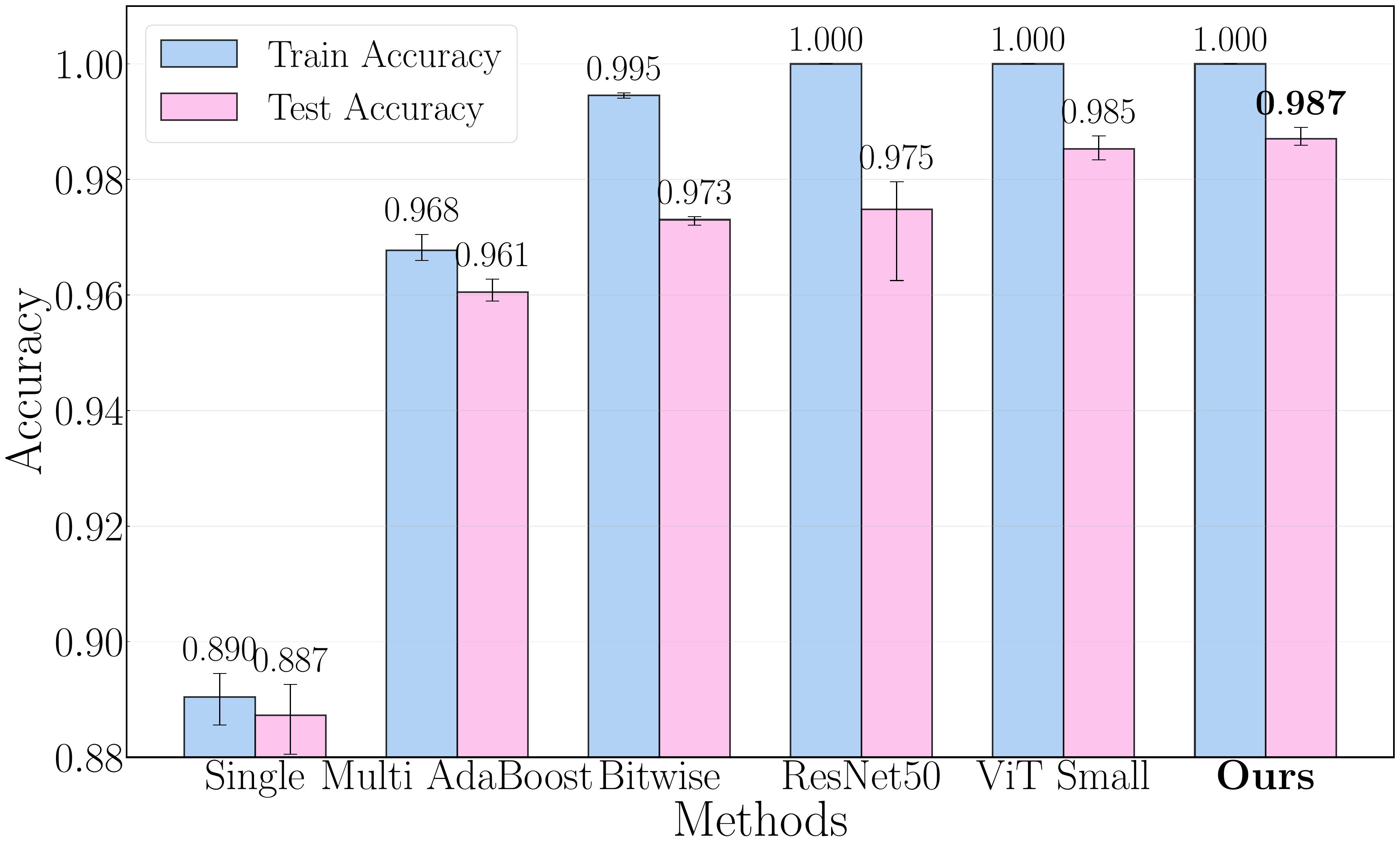}
    \caption{Training and test accuracy comparison on the selected MNIST subset between the TTA classifier and other approaches, including single quantum multi-class classifier, quantum multi-class AdaBoost classifier, quantum bitwise aggregate classifier, as well as classical neural networks such as ResNet50 classifier and ViT small classifier. The TTA classifier achieves the best training and prediction performance compared to all other methods. The numbers on the error bars represent the average values of 5 runs, and the error bars represent the maximum and minimum values across the 5 runs.}
    \label{fig:acc_compare}
\end{figure}

To validate the superiority of the TTA classifier, we conducted a comprehensive comparison with multiple quantum and classical classifiers on the selected MNIST subset, with results shown in Fig.~\ref{fig:acc_compare}.

\subsubsection{Accuracy Comparison}
We compared the TTA classifier with the following quantum methods: (1) a single quantum multi-class  classifier~\cite{schuld2020circuit},  (2) a multi-class AdaBoost classifier based on quantum multi-class models~\cite{wang2024supervised,li2024ensemble}, and (3) a bitwise aggregate classifier that uses bitwise encoding to transform the multi-class problem into multiple binary classification problems (with each binary classification problem using binary AdaBoost to enhance trainability). The experimental results demonstrate that while both the Multi AdaBoost ensemble classifier and bitwise aggregate classifier achieved improved training performance compared to the single quantum multi-class classifier, only the TTA classifier achieved 100\% train accuracy, thereby obtaining the best performance on test accuracy.

We also compared the TTA classifier with representative models from classical deep learning, including  (1) ResNet~\cite{he2016deep} (Residual Network, which effectively addresses the gradient vanishing problem in deep networks) and (2) ViT~\cite{dosovitskiy2020image} (Vision Transformer, a milestone model that successfully applies the Transformer architecture to computer vision tasks). Although these classical classifiers can also achieve 100\% train accuracy, benefiting from the excellent generalization properties of QML, the TTA classifier demonstrates the best performance on test accuracy.

\subsubsection{Efficiency Comparison}
\begin{figure}[H]
    \centering
    \includegraphics[width=0.48\textwidth]{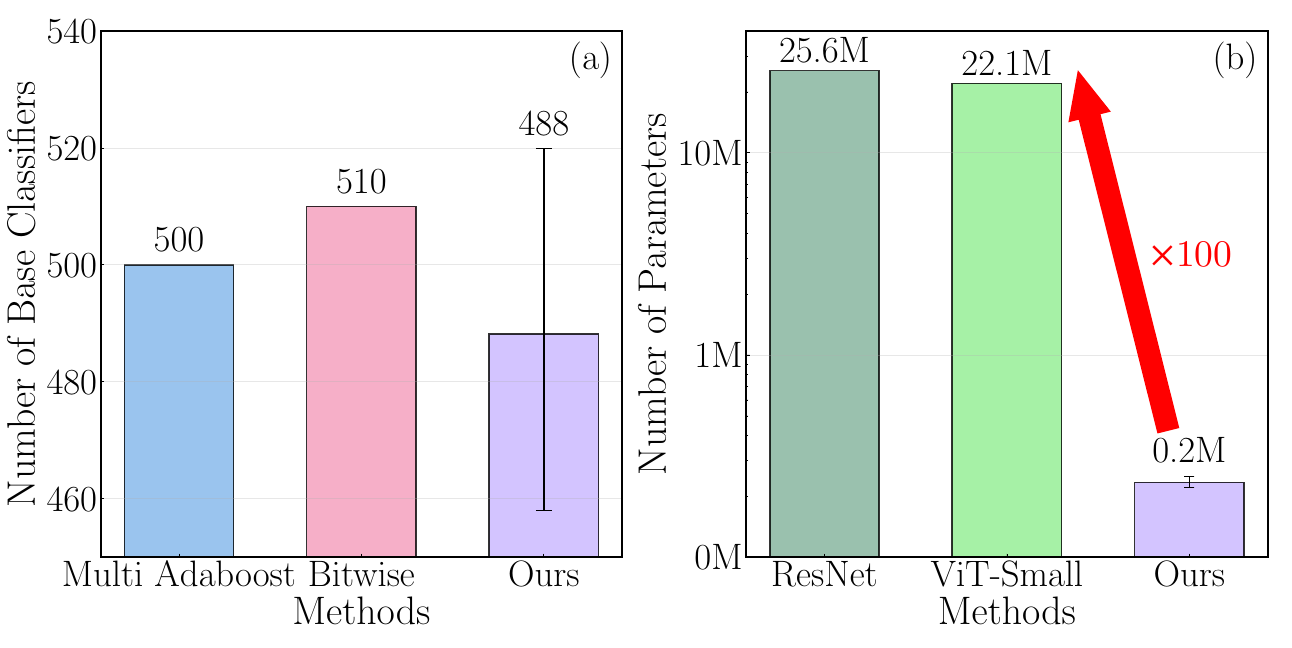}
    \caption{(a) Comparison of the number of base classifiers used by different quantum methods. (b) Parameter count comparison between the TTA classifier and classical neural networks. The numbers on the error bars represent the average values of 5 runs, and the error bars represent the maximum and minimum values.}
    \label{fig:number_params}
\end{figure}

To further evaluate the efficiency of TTA we conducted a comparative analysis from two dimensions: (1) relative to quantum classifiers, we compared the total number of AdaBoost base classifiers used by TTA versus other quantum methods; (2) relative to classical algorithms, we compared the parameter count of TTA versus classical classifiers. The results in Fig.~\ref{fig:number_params} demonstrate that compared to bitwise aggregate classifier and multi-class AdaBoost classifier, TTA uses the fewest base classifiers on average, while requiring approximately two orders of magnitude fewer parameters than classical models. Concretely, the TTA classifier uses approximately 500 base classifiers in total, with each base classifier containing 480 parameters (20 layers $\times$ 24 parameters per layer), resulting in approximately 0.2M total parameters. However, directly training an 8-qubit quantum circuit with 10,000 layers would be infeasible in practice due to barren-plateau and noise effects, which further highlights the importance and advantages of distributing capacity via the trace distance binary tree construction.

\subsection{Details on TTA classifier}\label{subsec:details_tta}

\subsubsection{Accuracy Approaches 100\% as Base Classifiers Increase}
\begin{figure}[htbp]
    \centering
    \includegraphics[width=0.49\textwidth]{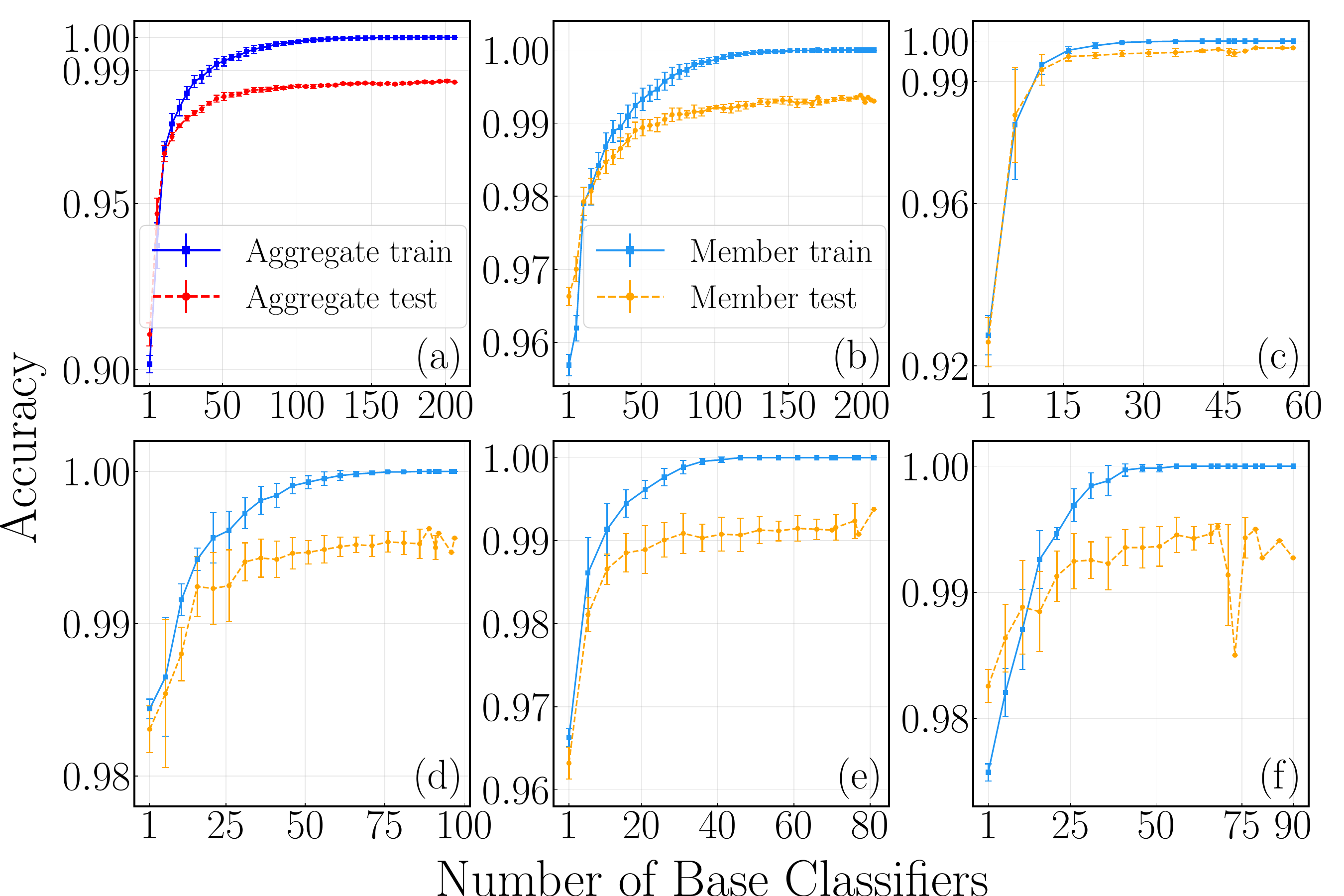}
    \caption{(a) Training and test accuracy of the aggregate classifier approach 1 as the number of base classifiers increases. (b)-(f) Training and test accuracy of each member classifier approach 1 as the number of base classifiers increases. 
    The lines represent the average results of 5 runs, and the error bars represent the maximum and minimum values across the 5 runs.}
    \label{fig:mnist_progress}
\end{figure}

We analyzed the changes in train and test accuracy for each member classifier and for the aggregate classifier as the number of base classifiers increases in Fig.~\ref{fig:mnist_progress}. Due to the characteristics of the AdaBoost algorithm, as the number of base classifiers increases, the train error of each member classifier decreases exponentially and the train accuracy approaches 1. This, in turn, causes the aggregate classifier's train accuracy to grow exponentially and reach 1. Benefiting from the superior generalization performance of the underlying QML models and from  AdaBoost's ability to maintain generalization properties, the test accuracy consistently remains close to the train accuracy, and when the train accuracy reaches 1, the test accuracy is also very close to 1.

In the experimental design, we measure train and test accuracy for each member classifier starting from 1 base classifier, then at every 5 additional base classifiers (i.e., at 1, 6, 11, 16, ...). Additionally, we perform one final measurement of training and test accuracy when each member classifier converges. Since all experiments are repeated 5 times with independent training, the number of base classifiers required for convergence varies across runs due to differences in base classifier errors. For the aggregate classifier, we aggregate results from all member classifiers at these fixed measurement points. When a member classifier converges before others, its final result is used for subsequent aggregate measurements. For example, if member classifier A converges at 21 base classifiers while member classifier B converges at 36, the aggregate results at measurement point 26 would use A's final result (at 21) combined with B's result at 26.

\subsubsection{Weak Base Classifiers Yield Strong Ensemble}
\begin{figure}[H]
    \centering
    \includegraphics[width=0.45\textwidth]{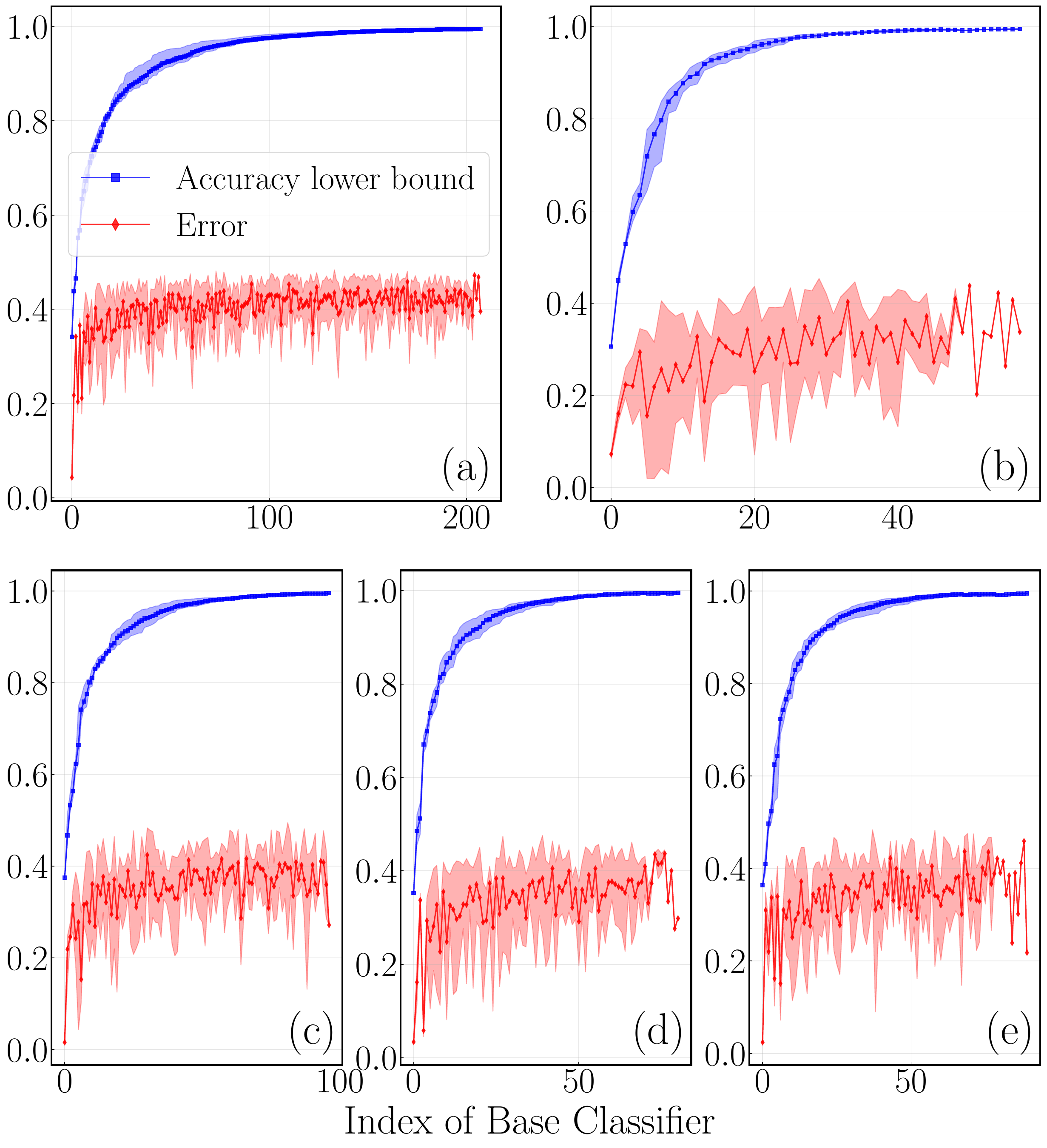}
    \caption{(a)-(e) Corresponding to the five member classifiers in Fig.~\ref{fig:mnist_progress}(b)-(f), showing the error of each base classifier and the accuracy lower bound of the entire AdaBoost member classifier up to that base classifier. The solid lines represent the average results of 5 independent runs, and the shaded areas represent the maximum and minimum values across the 5 independent runs.}
    \label{fig:mnist_bounds}
\end{figure}
We further analyze the error of each base classifier and the train accuracy lower bound of the AdaBoost member classifier in Fig.~\ref{fig:mnist_bounds}. The red curves show that individual base classifiers maintain a stable error of around 0.4 on the reweighted training set. Despite this, the train accuracy lower bound (computed as 1 minus the error upper bound guaranteed by AdaBoost) converges exponentially toward 1. All member classifiers successfully reached the convergence condition of train accuracy lower bound exceeding 0.995.

\subsubsection{Trace Distance and Sample Size Affect Convergence}
\begin{figure}[H]
    \centering
    \includegraphics[width=0.48\textwidth]{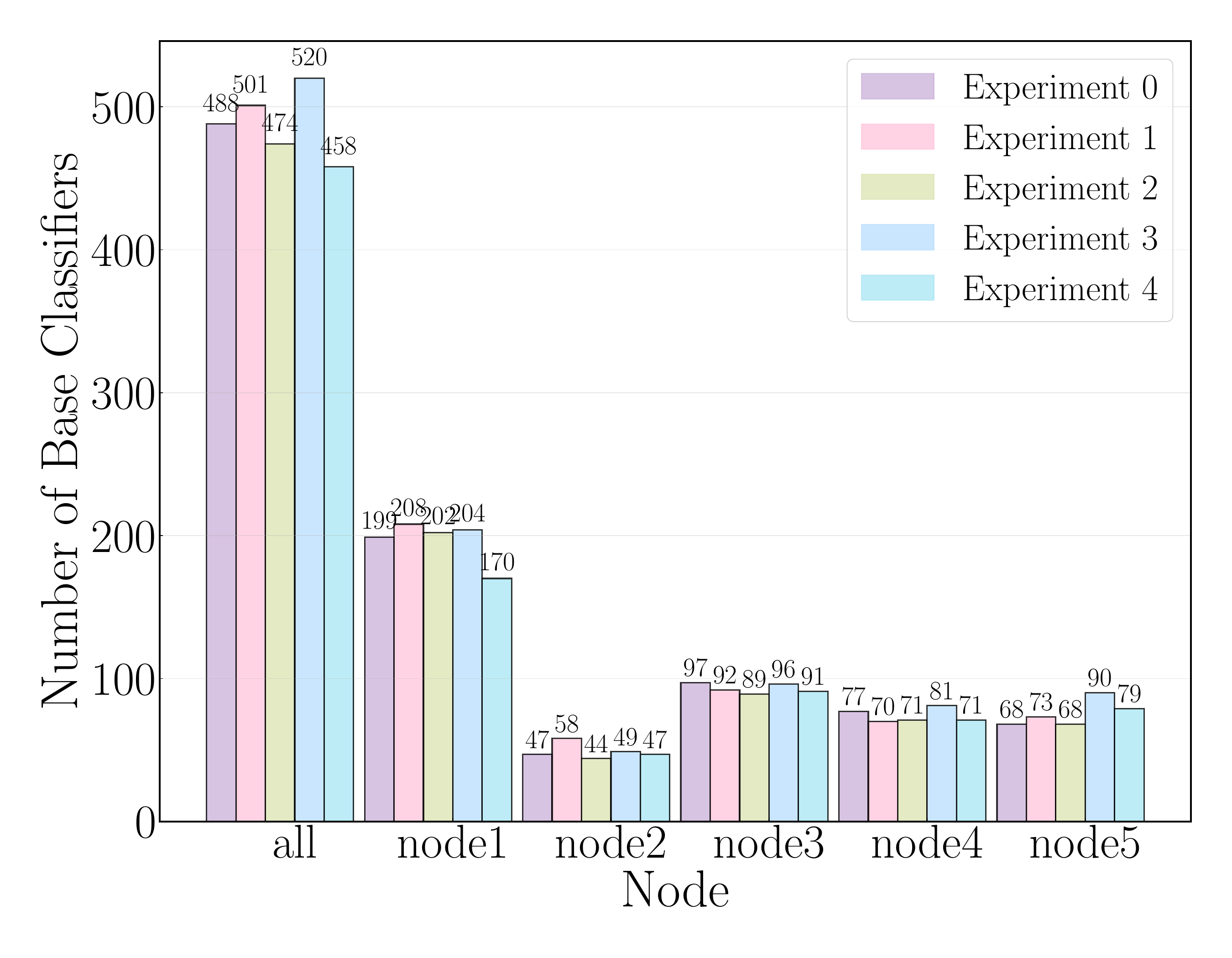}
    \caption{Number of base classifiers required for 5 nodes (member classifiers) to achieve an error upper bound less than 0.005 across 5 independent runs.}
    \label{fig:mnist_requiredN}
\end{figure}

Fig.~\ref{fig:mnist_requiredN} shows the number of AdaBoost base classifiers required for each member classifier to reach convergence across 5 independent runs. As shown in Table~\ref{tab:mnist_trace_distances}, the root node (node 1) has the smallest trace distance (0.5672) and the largest training set (12,000 samples), resulting in the most base classifiers required for convergence. In contrast, node 3 has the largest trace distance (0.7366), making the classification task easier and requiring fewer base classifiers. Comparing node 1 and node 5, although their trace distances are similar (0.5672 vs.\ 0.5593), node 5 requires significantly fewer base classifiers due to its smaller training set (4,000 vs.\ 12,000 samples). These results validate that both trace distance and sample size affect convergence speed, supporting our trace distance-based binary tree construction strategy.

\subsection{Details on Compared Methods}\label{subsec:details_compared}

\subsubsection{Bitwise Aggregate Classifier}

The bitwise aggregate method assigns each class a unique binary code and trains one classifier per bit position, thereby reducing multi-class classification to binary classification. For a $K$-class problem, the class labels are encoded as binary numbers of length $\left\lceil \log_2(K) \right\rceil$. A separate binary classifier is then trained for each bit position, where samples with 0 in that bit position form the negative class and those with 1 form the positive class.  To ensure a fair comparison with the TTA classifier, we also employ binary AdaBoost for each member classifier, constructing an aggregate classifier with enhanced trainability.

\begin{table}[h]
    \centering
    \caption{Trace distances between the average quantum states of the negative class (left) and positive class (right) for each member classifier in the bitwise aggregate classifier.}
    \label{tab:bitwise_trace_distances}
    \begin{tabular}{cccc}
    \hline
    \hline
    Bit & Classes  & Trace Distance & Training Samples \\
    \hline
    Bit 1 & \{0,1,3\} vs \{4,6,7\} & 0.5220 & 12000 \\
    Bit 2 & \{0,1,4\} vs \{3,6,7\} & 0.4184 & 12000 \\
    Bit 3 & \{0,4,6\} vs \{1,3,7\} & 0.5500 & 12000 \\
    \hline
    \hline
    \end{tabular}
    \end{table}

Specifically, in the selected MNIST subset, since the largest label is 7, binary strings of length 3 are required to encode the labels in the dataset, converting the 6-class problem into 3 binary classification problems. Taking label 6 as an example, its binary representation is 110. Following the bit order from most (left) to least significant (right), label 6 belongs to the positive class in the first and second member classifiers, and to the negative class in the third member classifier. The trace distances between the positive and negative class average states for the three member classifiers are shown in Table~\ref{tab:bitwise_trace_distances}. As can be seen, under these three partitioning scenarios, the trace distances of the average states are all smaller than those in the TTA partitioning approach, indicating that the binary classification tasks are, on average, more difficult.

\begin{figure}[H]
    \centering
    \includegraphics[width=0.45\textwidth]{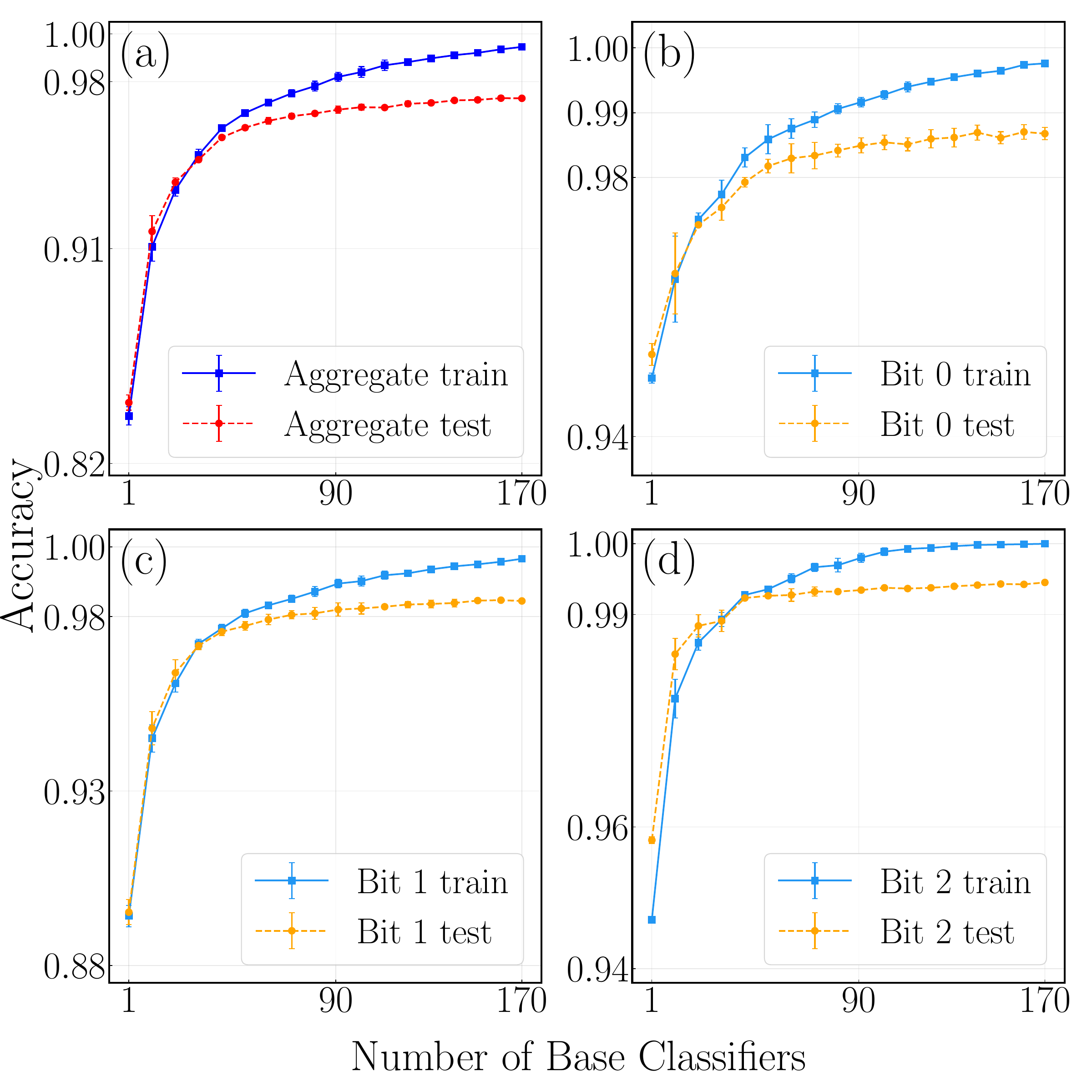}
    \caption{(a) Training and test accuracy of the bitwise aggregate classifier. (b)-(d) Training and test accuracy of each bitwise member classifier.
    The lines represent the average results of 5 independent runs, and the error bars represent the maximum and minimum values across the 5 independent runs.}
    \label{fig:bitwise}
\end{figure}

For fair comparison, we configure the bitwise aggregate to use a similar number of base classifiers as our TTA classifier, which employs an average of 488 base classifiers. We set the maximum number of AdaBoost rounds for each member classifier to 170, resulting in a total of 510 base classifiers for the bitwise aggregate—slightly more than the TTA classifier.

However, it is worth noting that each base classifier in the bitwise aggregate must be trained on the complete set of 12,000 samples, whereas the TTA classifier only trains node 1 on 12,000 samples while nodes 2--5 are trained on smaller subsets ranging from 4,000 to 6,000 samples (see Table~\ref{tab:mnist_trace_distances}). This difference results in significantly higher training difficulty for the bitwise aggregate. All other training settings remain identical to those of the TTA classifier. As shown in Fig.~\ref{fig:bitwise}, although training performance improves as the number of base classifiers increases, the member classifiers struggle to achieve high train accuracy (100\%). This is because each member classifier must process 12,000 samples without reasonable partitioning based on trace distance. Consequently, despite underlying binary models can generalize well in principle, bitwise aggregates' test performance remains poor due to these training difficulties.

\subsubsection{Single Quantum Multi-Class Classifier}

For the single quantum multi-class classifier trained on the selected MNIST subset, we design measurement operators $\Pi_{k} = \ket{k}\bra{k}$ for each class $k$ to handle multi-class classification. The model parameters are optimized using cross-entropy loss on the dataset $S = \{(\rho^{(m)},y^{(m)})\}_{m=1}^{M}$, with the loss function defined as:
\begin{equation}
    \label{eq:cross_entropy_loss}
\begin{aligned}
    &\mathcal{L}(S;\boldsymbol{\theta}) = \frac{1}{M} \sum_{m=1}^{M} \ell(\rho^{(m)},y^{(m)};\boldsymbol{\theta}), \\
    &\ell(\rho^{(m)},y^{(m)};\boldsymbol{\theta}) = -\sum_{k=1}^{K}  \boldsymbol{y}_k^{(m)} \ln \left( \frac{e^{h_k(\rho^{(m)},\boldsymbol{\theta})}}{\sum_{j=1}^{K} e^{h_j(\rho^{(m)},\boldsymbol{\theta})}} \right), \\
    &h_k(\rho^{(m)},\boldsymbol{\theta}) = \operatorname{Tr}\left[\Pi_k U(\boldsymbol{\theta}) \rho^{(m)} U^{\dagger}(\boldsymbol{\theta})\right],
\end{aligned}
\end{equation}
where $\boldsymbol{y}_k^{(m)}$ represents the $k$-th element of the one-hot encoded vector corresponding to the true label $y^{(m)}$ of sample $m$. Specifically, when $y^{(m)} = l$, meaning that sample $m$ belongs to class $l$, we have $\boldsymbol{y}_l^{(m)} = 1$ and all other positions are 0.

\begin{figure}
    \centering
    \includegraphics[width=0.45\textwidth]{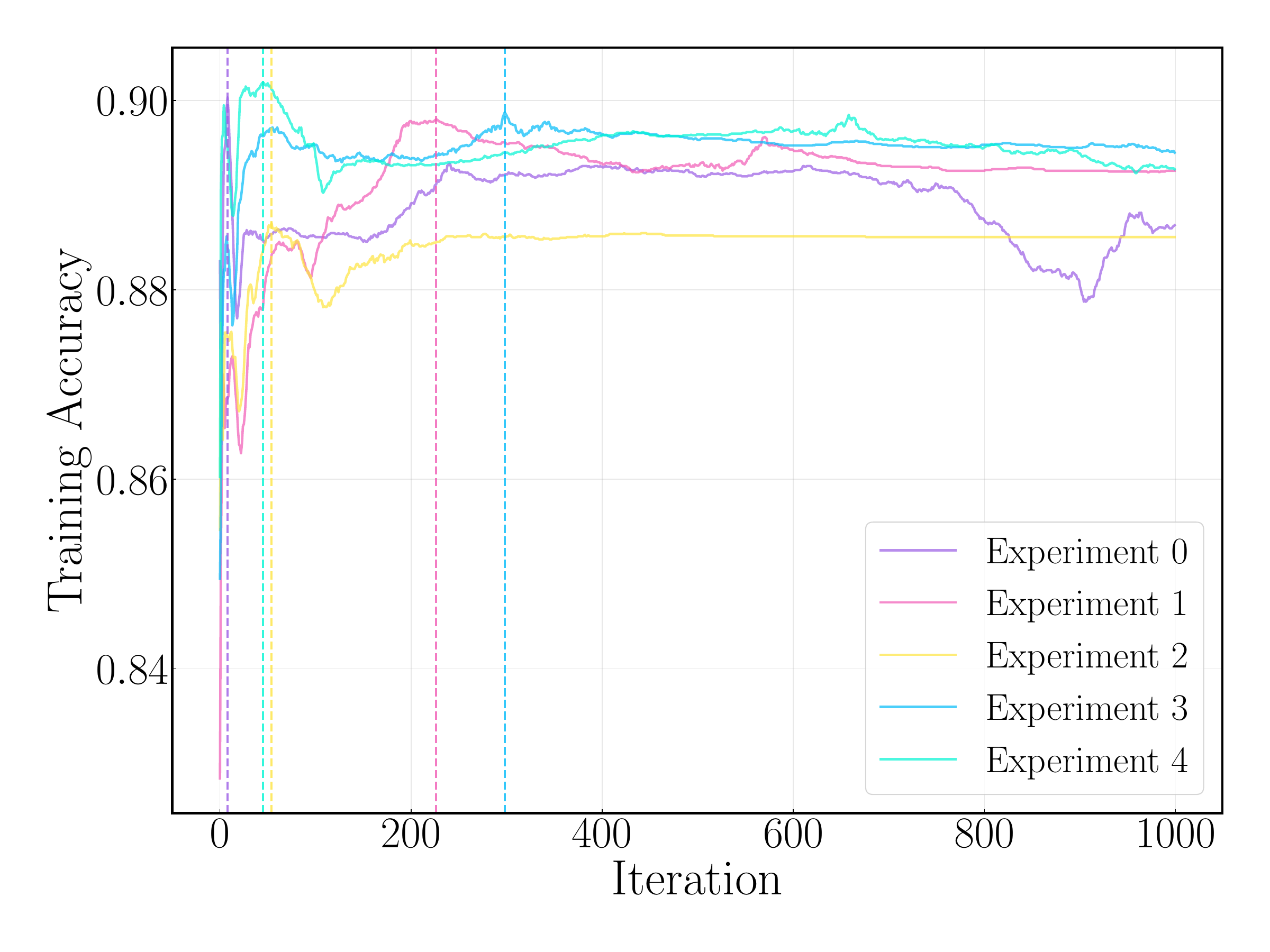}
    \caption{Train accuracy of a single QML model for multi-class classification. The vertical lines indicate the results with the highest train accuracy during the 1000 training epochs.}
    \label{fig:single_model}
\end{figure}

To investigate whether simply increasing the number of training epochs would significantly improve the train accuracy of a single quantum multi-class classifier, we set the number of training epochs to 1000 and disabled early stopping. All other settings, including optimizer, batch size, and learning rate, remain consistent with the TTA classifier to ensure fair experimental comparison. The experimental results are shown in Fig.~\ref{fig:single_model}, where we can observe that across 5 independent runs, even with 1000 training epochs, the epoch achieving the best train accuracy occurs within the first 400 epochs, indicating that simply increasing training iterations cannot significantly improve the train accuracy of a single quantum multi-class classifier.

Additionally, since the TTA classifier uses approximately 10,000 quantum circuit layers in total, and excessive expressivity leads to barren plateaus, training a single multi-class QML classifier with 10,000 layers would be impractical.

\subsubsection{Multi-class AdaBoost Classifier}

Previous work~\cite{li2024ensemble,wang2024supervised} has introduced multi-class AdaBoost into QML to improve trainability. However, unlike binary AdaBoost, multi-class AdaBoost methods do not theoretically guarantee that as long as each base classifier achieves an error rate below a certain threshold on the reweighted training set, the overall ensemble AdaBoost train error will gradually converge to 0. Therefore, simply increasing the number of base classifiers in multi-class AdaBoost does not necessarily achieve the 100\% train accuracy that the TTA classifier achieves.

When training the $t$-th multi-class base classifier, we adopt the cross-entropy loss function on the weighted training set $S = \{(\rho^{(m)},y^{(m)})\}_{m=1}^{M}$ as follows:
$$
\begin{aligned}
L(S,\boldsymbol{\theta}) = \sum_{m=1}^{M} \boldsymbol{w}_t(m) \ell(\rho^{(m)},y^{(m)},\boldsymbol{\theta})
\end{aligned}
$$
Here, $\boldsymbol{w}_t(m)$ is the weight assigned by multi-class AdaBoost to the $m$-th sample when training the $t$-th base classifier, and $\ell(\rho^{(m)},y^{(m)},\boldsymbol{\theta})$ has the same definition as in Eq.~\eqref{eq:cross_entropy_loss}.

Similarly, to ensure fair comparison, we configure the multi-class AdaBoost to use a similar number of base classifiers as our TTA classifier, which employs an average of 488 base classifiers. We set the maximum number of base classifiers for the entire multi-AdaBoost to 500, which exceeds the average number of base classifiers used by the TTA classifier. Additionally, we modify the early stopping strategy as follows: when the train error of a base classifier is less than $5/6$ and no decrease in train error occurs for 10 consecutive epochs, training is stopped. This is because multi-class AdaBoost only requires the train error of base classifiers to be less than $(K-1)/K$ (see App.~\ref{app:other_algorithms} for details). All other settings, including optimizer, batch size, and learning rate, remain consistent with the TTA classifier to ensure fair experimental comparison.
\begin{figure}[H]
    \centering
    \includegraphics[width=0.45\textwidth]{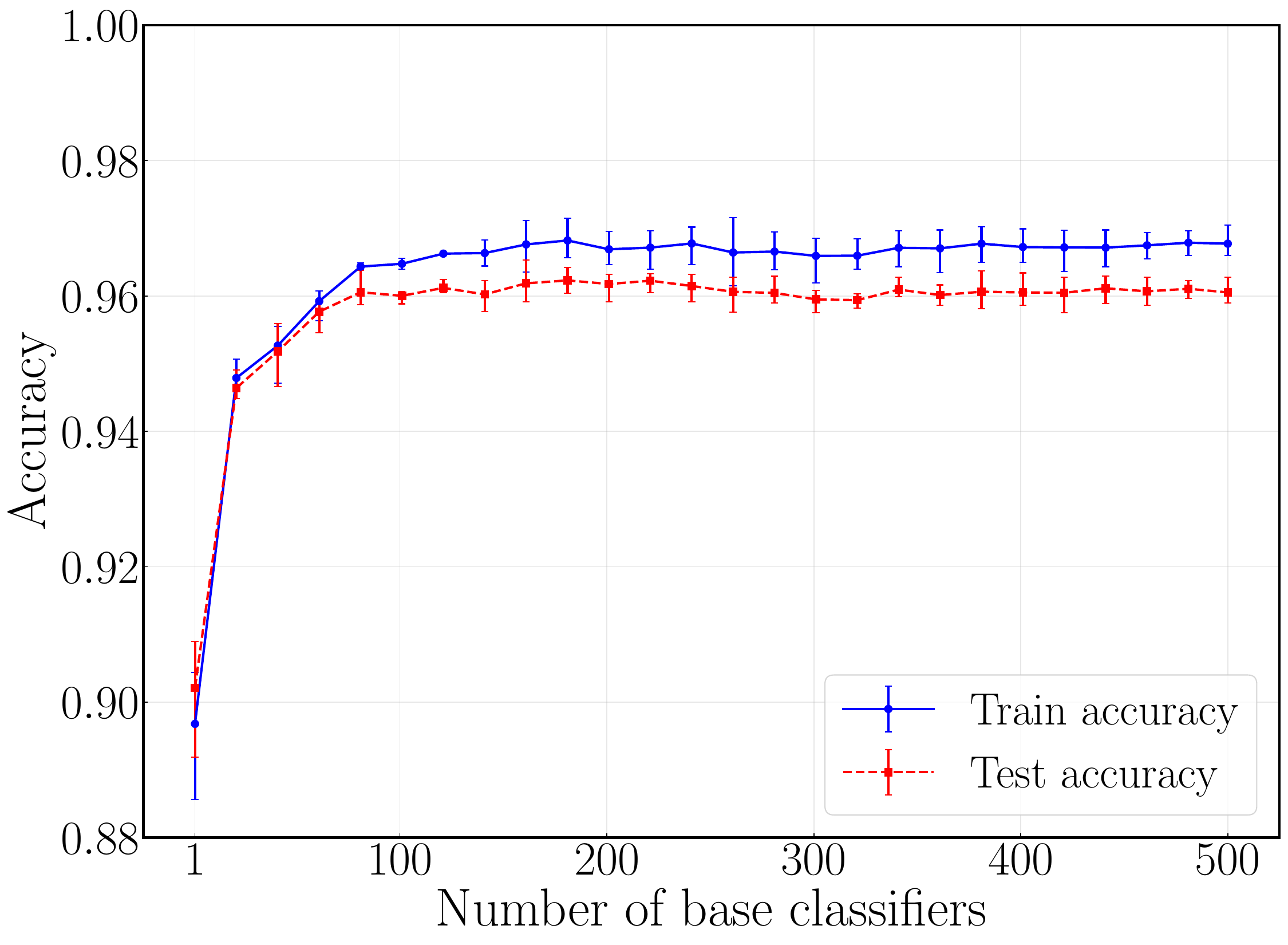}
    \caption{Train and test results using multi-class AdaBoost. It can be observed that the train accuracy of each classifier fails to reach 1, while the test accuracy remains close to the train accuracy. The lines represent the average results of 5 independent runs, and the error bars represent the maximum and minimum values across the 5 independent runs.}
    \label{fig:multiadaboost}
\end{figure}

The experimental results are shown in Fig.~\ref{fig:multiadaboost}. Train accuracy initially improves with more base classifiers but plateaus beyond a certain threshold, while test accuracy remains close to train accuracy, confirming the strong generalization of QML models. This limitation stems from the lack of theoretical guarantees in multi-class AdaBoost: unlike binary AdaBoost, where train error decreases exponentially with the number of base classifiers, multi-class AdaBoost offers no such guarantee, preventing further trainability improvements by simply adding more base classifiers.

\subsubsection{Classical Neural Network}
\begin{figure}[htpb]
    \centering
    \includegraphics[width=0.43\textwidth]{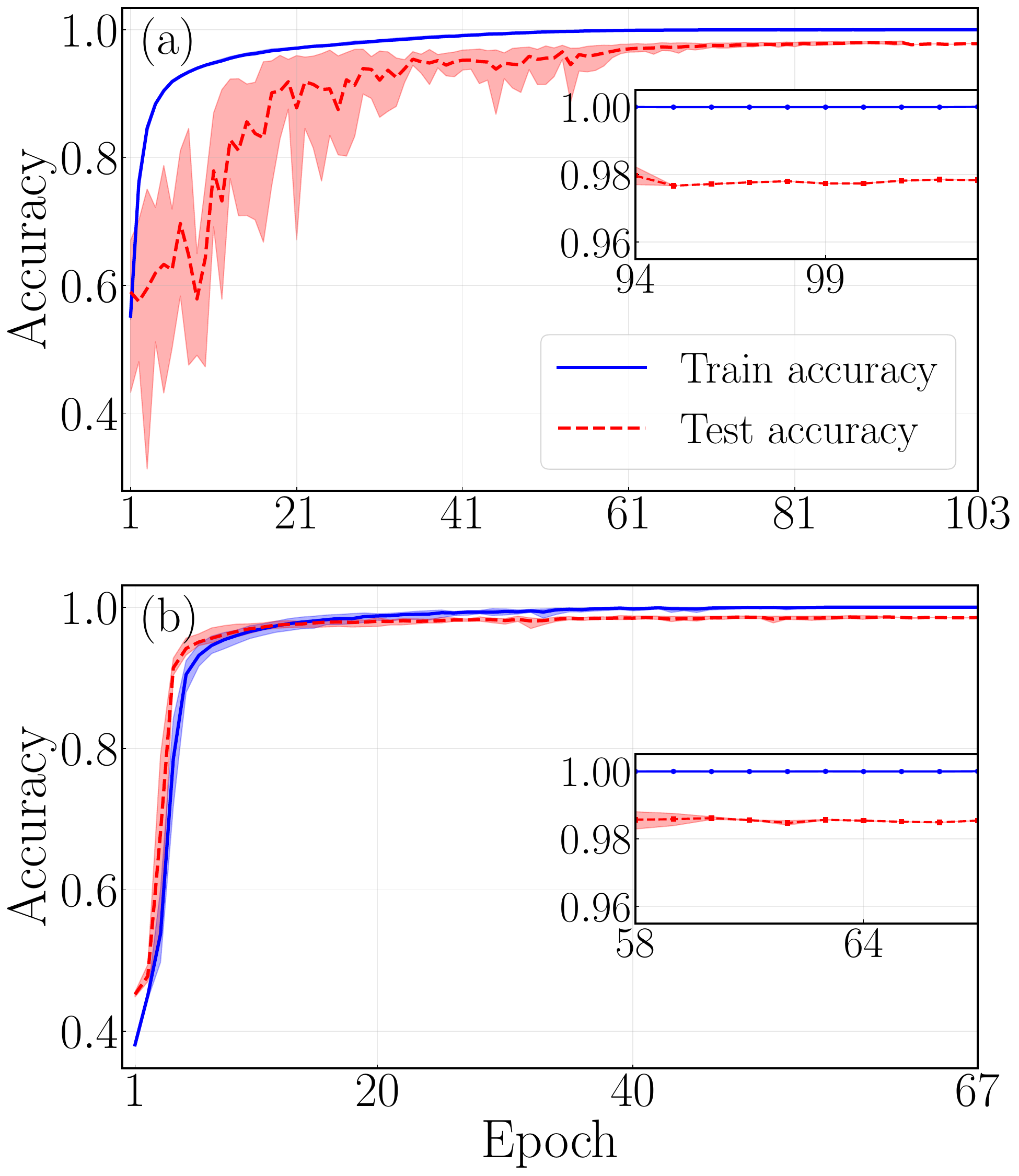}
    \caption{(a) ResNet and (b) ViT training and test results. The lines represent the average results of 5 independent runs, and the shadows represent the maximum and minimum values across the 5 independent runs. The insets show the last 10 training epochs.}
    \label{fig:MNIST_classical}
\end{figure}
To evaluate the TTA classifier, we compare it with two classical deep learning models: ResNet and Vision Transformer (ViT). Since our dataset consists of $16 \times 16$ single-channel images, we first interpolate them to $224 \times 224$ images, which is a standard practice in classical deep learning. Then, we only need to modify the input channel number of both models. For ResNet, we changed the input channel number of the first convolutional layer from 3 (the default for RGB images) to 1, while keeping all other architectural parameters unchanged. For ViT, we similarly modified the patch embedding layer to accept single-channel input instead of the default 3-channel RGB input, with all other architectural parameters unchanged.

During the training process, we adopt the cross-entropy loss function as the optimization objective and select the SGD optimizer~\cite{zhou2020towards} for parameter updates, primarily because SGD often achieves superior model generalization performance in practice. Both neural networks are trained from scratch without using pre-trained weights. For training configuration, we set the maximum number of training epochs to 1000 and establish a clear early stopping strategy: once the model achieves 100\% accuracy on the training set, the training process is terminated. To ensure fair experimental comparison, other hyperparameter settings (such as batch size, learning rate, etc.) remain completely consistent with the TTA classifier. All experiments are repeated 5 times randomly.

The experimental results are shown in Fig.~\ref{fig:MNIST_classical}. From the results, we can observe that although classical neural networks can relatively easily achieve 100\% accuracy on the training set, due to their generalization capability being somewhat inferior compared to QML models, resulting in final classification performance that is slightly inferior to our proposed TTA classifier. Since the classifiers stop training once they achieve 100\% accuracy on the dataset, the plots are aligned based on the maximum number of epochs from the experiments. Additionally, we explain in App.~\ref{app:sgd_reason} why we chose the SGD optimizer over the Adam optimizer; at the same learning rate, ViT performs worse with the Adam optimizer compared to SGD.

\section{Classification on Quantum Data}\label{sec:quantum}

This section demonstrates the effectiveness of the TTA classifier on quantum data. Subsec.~\ref{subsec:annni_model} introduces the ANNNI model and quantum phase classification task. Subsec.~\ref{subsec:quantum_performance} presents the performance results. Subsec.~\ref{subsec:quantum_visualization} provides visualization of training and test accuracy.

\subsection{ANNNI model}\label{subsec:annni_model}

We consider an ANNNI model~\cite{selke1988annni} with $N$ qubits, whose Hamiltonian is given by $$H=-\left(\sum_{i=1}^{N-1} X_{i} X_{i+1}-\kappa \sum_{i=1}^{N-2} X_{i} X_{i+2}+h \sum_{i=1}^{N} Z_{i}\right).$$ Depending on the values of $\kappa$ and $h$, the ground state of this Hamiltonian exhibits in three distinct quantum phases: (i) Antiphase (labeled as 0), (ii) Ferromagnetic phase (labeled as 1), and (iii) Paramagnetic phase (labeled as 2). The phase boundaries are determined by two critical lines: $h_I(\kappa) \approx  \frac{1-\kappa}{\kappa}\left(1-\sqrt{\frac{1-3 \kappa+4 \kappa^{2}}{1-\kappa}}\right)$ and $h_{C}(\kappa) \approx 1.05 \sqrt{(\kappa-0.5)(\kappa-0.1)}$. In our experiments, we use these phase labels to classify the ground states.

\begin{figure}
    \centering
    \includegraphics[width=0.5\textwidth]{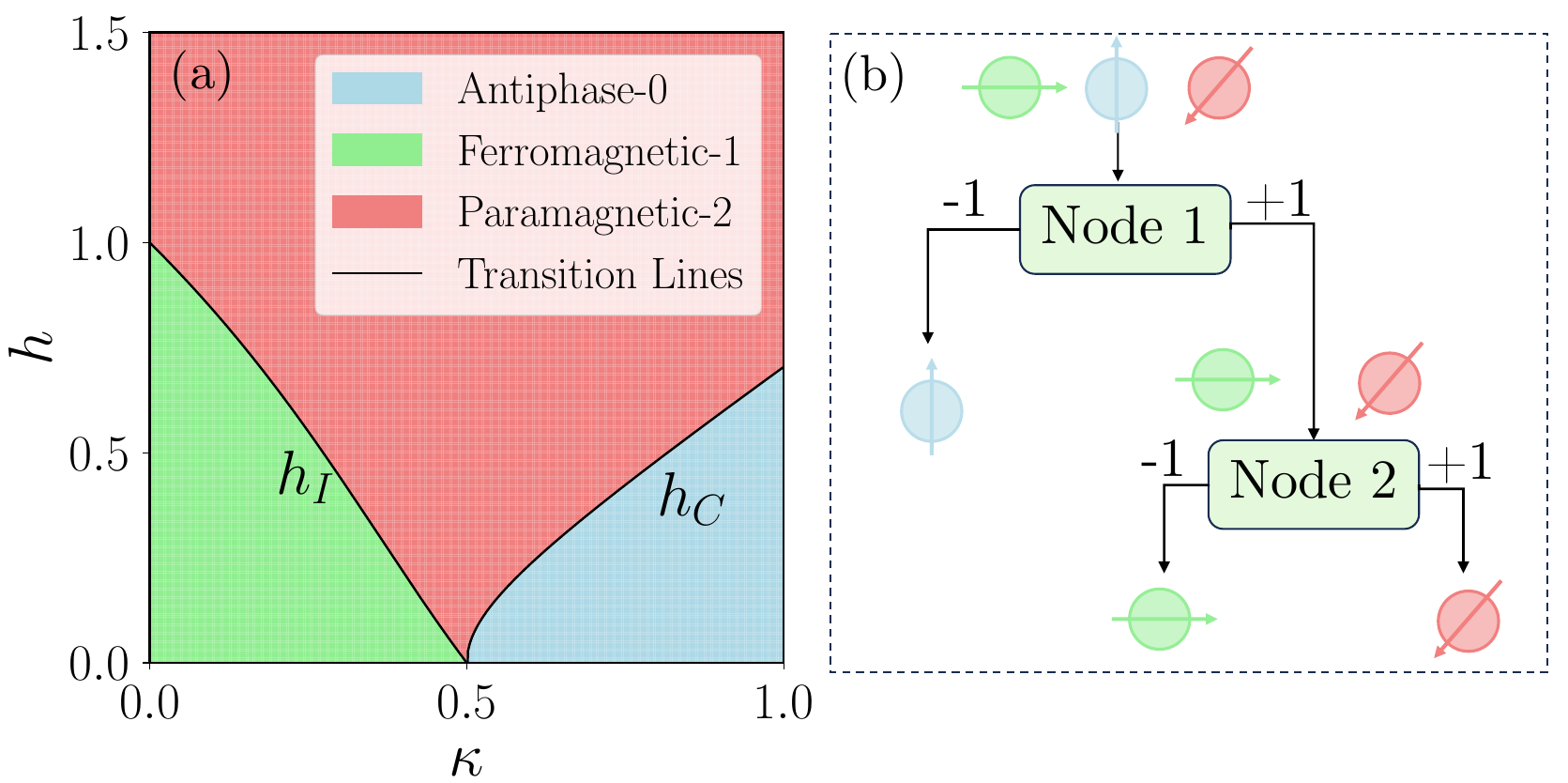}
    \caption{(a) Three phases of the ANNNI model (b) Trace distance tree for ANNNI phase dataset. For Node1, the Antiphase is treated as the negative class, while the Ferromagnetic and Paramagnetic phases are treated as the positive class. For Node2, the Ferromagnetic phase is treated as the negative class, and the Paramagnetic phase is treated as the positive class.}
    \label{fig:phase_tree}
\end{figure}

The 6-qubit phase diagram and the corresponding trace distance tree are shown in Fig.~\ref{fig:phase_tree}. We randomly selected 2000 quantum states for each phase as the training set and 1000 quantum states as the test set, resulting in a total of 6000 training samples and 3000 test samples. The class assignments for each node, along with the trace distances between positive and negative average states and the number of training samples involved, are summarized in Table~\ref{tab:phase_trace_distances}.

The training procedure for the TTA classifier on the ANNNI model follows the same protocol as described in Sec.~\ref{subsec:mnist_dataset} for classical data. Specifically, we use the PQC architecture shown in Fig.~\ref{fig:arch} with circuit depth $L=20$, the Pauli Z operator on the first qubit as the observable, hinge loss as the loss function, a batch size of 200, the Adam optimizer with a learning rate of 0.005, and the same early stopping strategy. All experiments are repeated 5 times.

\begin{table}[h]
    \centering
    \caption{Trace distances between the average quantum states of the negative class (left) and positive class (right) for each node in the phase TTA tree.}
    \label{tab:phase_trace_distances}
    \begin{tabular}{cccc}
    \hline
    \hline
    Node & Classes  & Trace Distance & Training Samples \\
    \hline
    Node 1 & \{0\} vs \{1,2\} & 0.8573 & 6000 \\
    Node 2 & \{1\} vs \{2\} & 0.7521 & 4000 \\
    \hline
    \hline
    \end{tabular}
    \end{table}
\subsection{Performance}\label{subsec:quantum_performance}

\subsubsection{Accuracy Approaches 100\% as Base Classifiers Increase}
\begin{figure}
    \centering
    \includegraphics[width=0.49\textwidth]{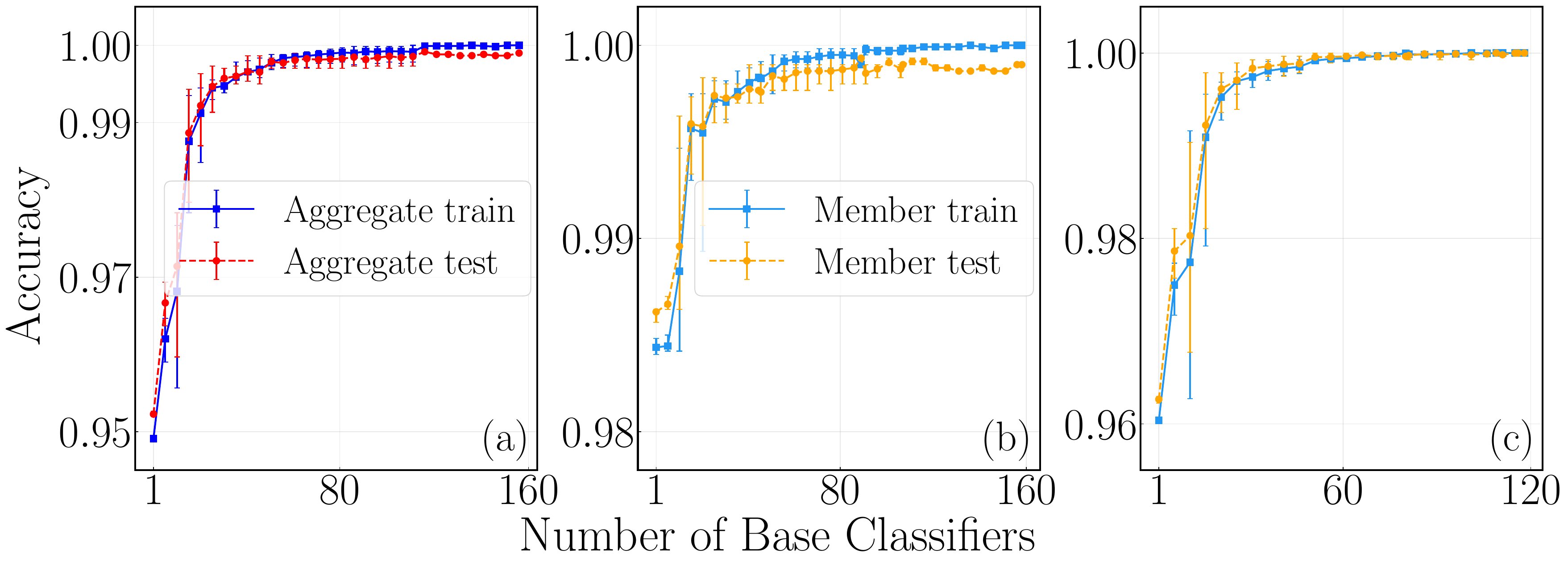}
    \caption{(a) As the number of base classifiers increases, both training and test accuracy approach 1. (b)-(c) Similar trend for individual member classifiers. The lines represent the average results of 5 independent runs, and the error bars represent the maximum and minimum values across the 5 independent runs.}
    \label{fig:phase_results}
\end{figure}

Following the same measurement strategy as in Sec.~\ref{subsec:details_tta}, we analyze the train and test accuracy as the number of base classifiers increases. As shown in Fig.~\ref{fig:phase_results}, both member classifiers and the aggregate classifier achieve train accuracy approaching 1, with test accuracy remaining close to train accuracy due to the strong generalization of QML models.

\subsubsection{Near-Perfect Classification on Phase Diagram}
\begin{figure}[H]
    \centering
    \includegraphics[width=0.48\textwidth]{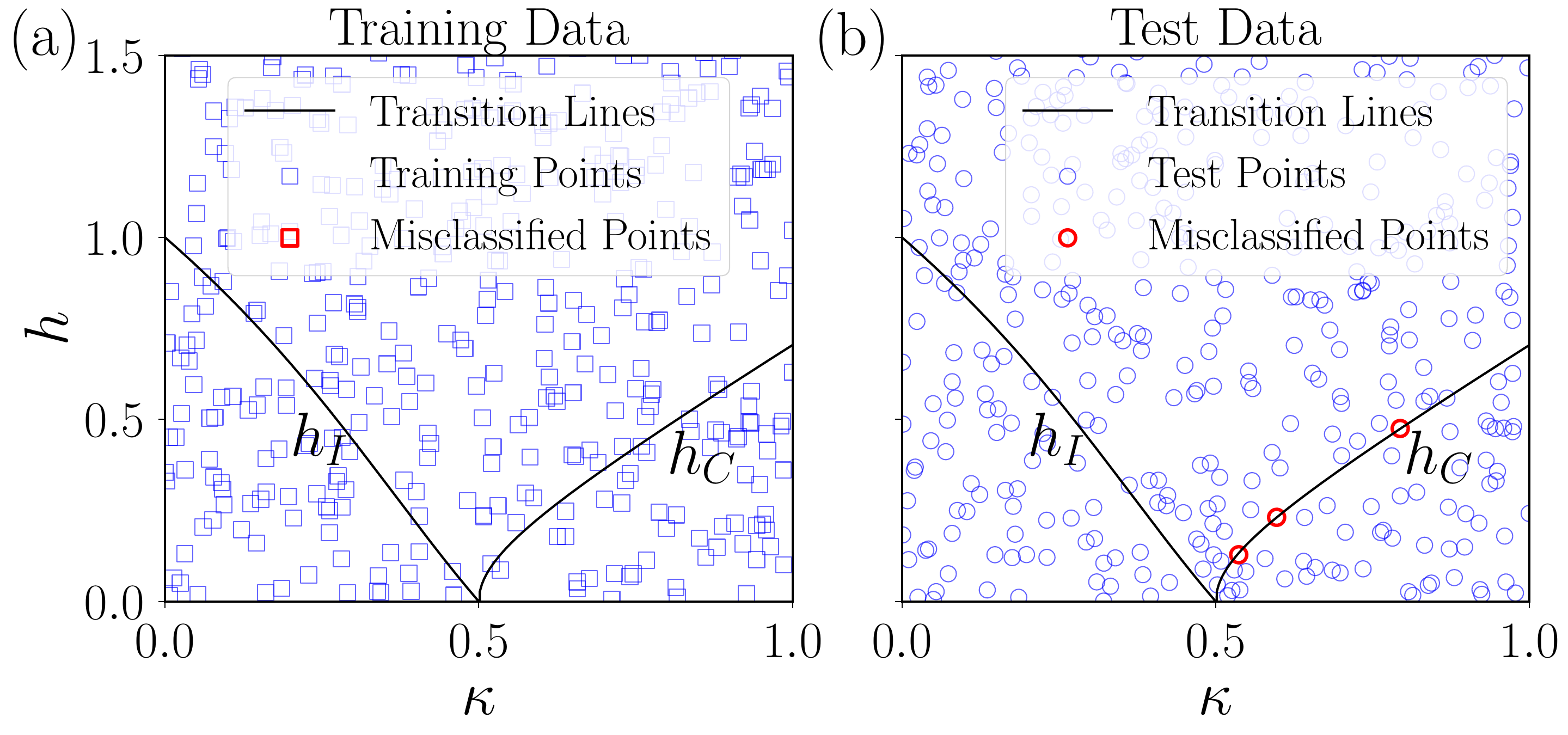}
    \caption{Training and prediction results of TTA on the quantum phase classification problem. Blue points represent correctly classified samples, while red points represent misclassified samples. In (a), we successfully classify all training data (6000 points) correctly. In (b), only 3 out of 3000 test samples are misclassified, and these errors are all concentrated near the classification decision boundary.}
    \label{fig:phase_scatter}
\end{figure}
To provide a more intuitive demonstration of classification performance, we select the experimental result with the smallest test error from 5 independent runs and visualize the classification results on the phase plane in Fig.~\ref{fig:phase_scatter}. The experimental results demonstrate that the TTA classifier achieves perfect classification on the training set, with all 6000 training samples correctly classified. On the test set, only 3 out of 3000 test samples are misclassified, and these misclassified samples are all located near the decision boundaries between different quantum phases. This phenomenon fully illustrates the effectiveness of the TTA classifier: (1) The perfect performance on the training set proves the effectiveness of the trace distance binary tree construction strategy and AdaBoost ensemble learning; (2) The high accuracy on the test set (99.9\%) demonstrates the excellent generalization capability of QML models; (3) The phenomenon that misclassified samples concentrate near decision boundaries aligns with the expectations of classification theory, indicating that the TTA classifier can accurately identify the essential differences between different quantum phases, with only minor uncertainties existing in the critical regions of phase transitions.

\subsection{Behavior of Sample Weights in AdaBoost}\label{subsec:quantum_visualization}

To analyze the learning mechanism of the AdaBoost algorithm in depth, we further investigate the dynamic changes in sample weights during the training process. The core idea of the AdaBoost algorithm lies in adaptively adjusting the weights of training samples to progressively improve the performance of the ensemble classifier. Specifically, in the initial stage of the algorithm, all training samples are assigned equal weights, reflecting fair treatment of each sample. As the iterative process proceeds, those samples that are misclassified by the current classifier receive higher weights, thus gaining more attention in the training of subsequent base classifiers. This weight redistribution mechanism enables the algorithm to gradually focus on those "difficult samples" that are hard to classify correctly, ultimately achieving significant improvement in overall classification performance.
\begin{figure}[H]
    \centering
    \includegraphics[width=0.48\textwidth]{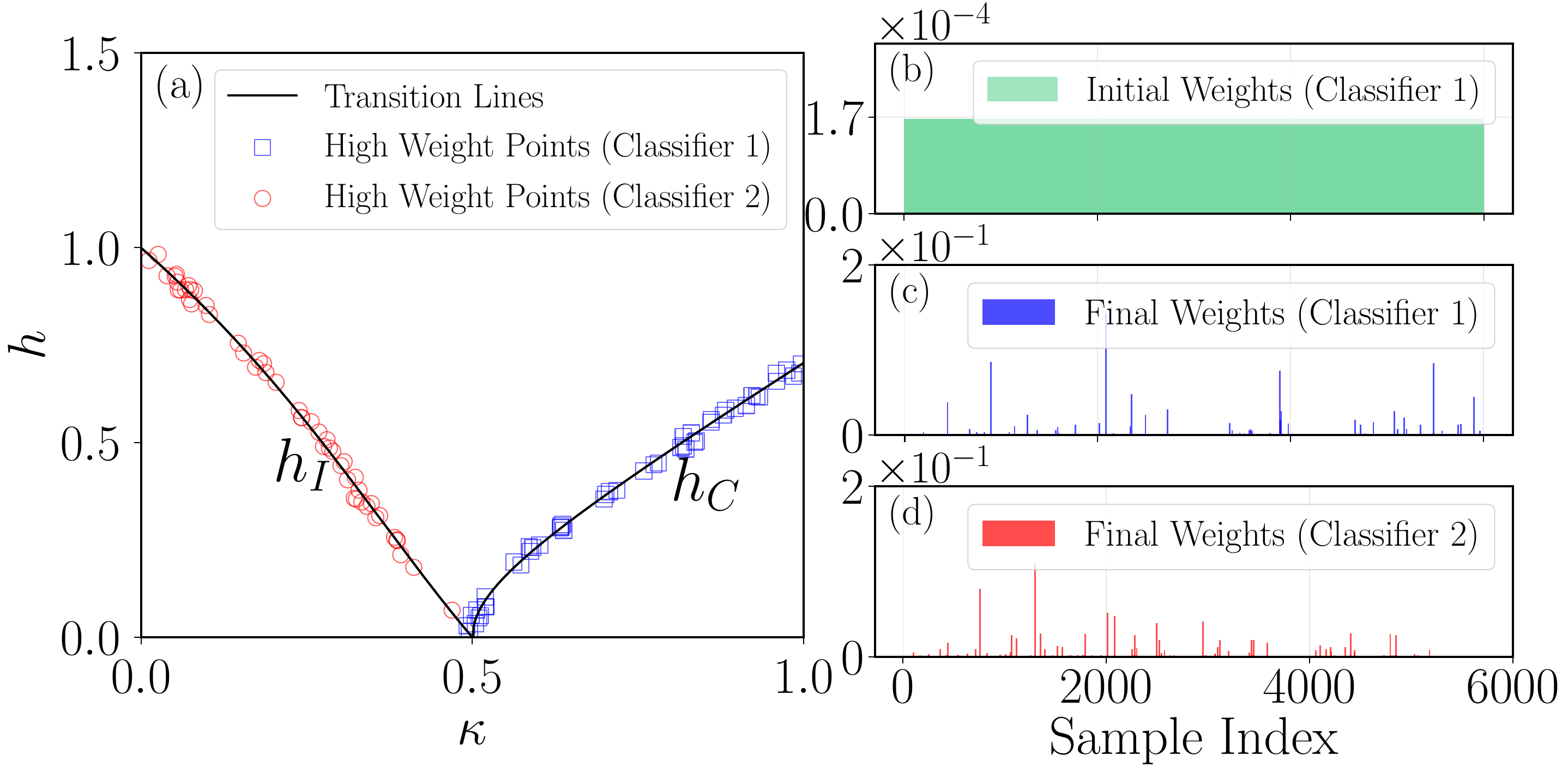}
    \caption{(a) High-weight training samples for both member classifiers are concentrated near decision boundaries. (b) Initial uniform weight distribution for member classifier 1 (the initial distribution for member classifier 2 is also uniform). (c)-(d) Final weight distributions for member classifier 1 and member classifier 2, respectively.}
    \label{fig:weights}
\end{figure}

In the ANNNI quantum phase classification experiment, the TTA classifier contains two member classifiers corresponding to the two nodes in the trace distance tree shown in Fig.~\ref{fig:phase_tree}(b). Using the same experimental result visualized in Fig.~\ref{fig:phase_scatter}, we extract the sample weight distribution corresponding to the final base classifier when each member classifier converges, and mark the training samples with higher weights in the $\kappa$-$h$ phase plane in Fig.~\ref{fig:weights}.

The experimental results clearly demonstrate the intelligent learning characteristics of the AdaBoost algorithm: the training samples with the highest weights (i.e., the most difficult samples to classify) are mainly concentrated near the decision boundaries between different quantum phases. This phenomenon has profound physical significance and a solid algorithmic theoretical foundation: (1) From a physics perspective, quantum states located near phase transition boundaries often possess more complex entanglement structures and more subtle quantum correlations, making them exhibit higher difficulty in classification tasks; (2) From a machine learning perspective, samples near decision boundaries naturally have higher classification uncertainty, requiring classifiers to invest more "attention" to accurately identify them.

\section{Classification on Synthetic Data}\label{sec:robustness}

This section systematically tests the robustness of the TTA classifier under various noise environments. Subsec.~\ref{subsec:synthetic_dataset} introduces the synthetic classification dataset. Subsec.~\ref{subsec:performance_noisy} presents performance results under different noise types. Subsec.~\ref{subsec:comparison_ovo_ovr} compares the TTA classifier with OVO and OVR approaches.

\subsection{Synthetic Classification Dataset}\label{subsec:synthetic_dataset}

\begin{figure}
    \centering
    \includegraphics[width=0.5\textwidth]{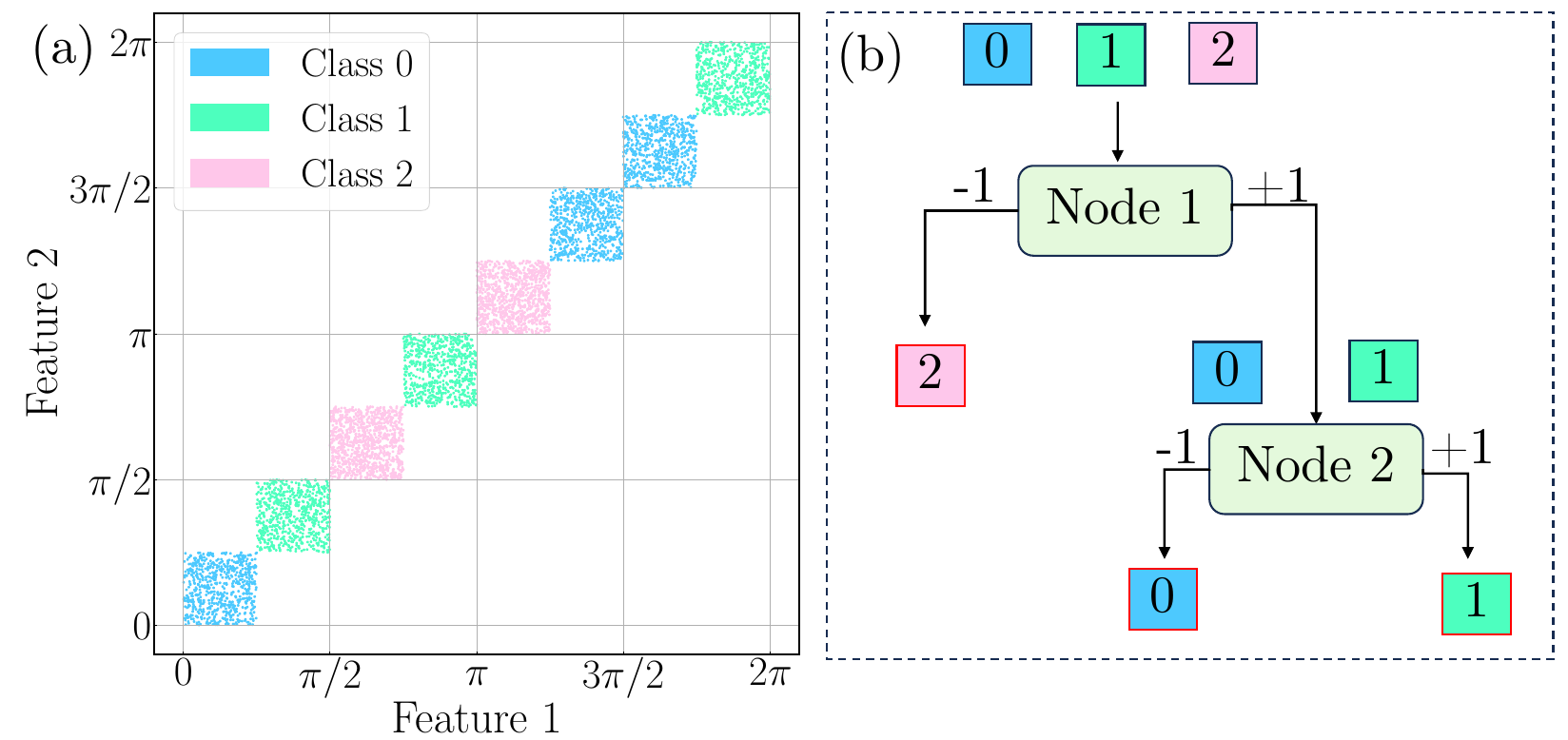}
    \caption{(a) Synthetic dataset visualization. (b) Trace distance tree for synthetic dataset.}
    \label{fig:synthetic}
\end{figure}

To systematically evaluate the impact of quantum noise on the TTA classifier, we construct a synthetic dataset. Specifically, we uniformly divide the interval $[0,2\pi]$ into 8 equal-length subintervals, and within each subinterval (such as $[0,\pi/4]$, $[\pi/4,\pi/2]$, etc.), we randomly generate $d$-dimensional vector data according to a uniform distribution. Subsequently, we randomly assign the $d$-dimensional data points generated from these 8 different subintervals to three classes (labeled as 0, 1, 2), thereby constructing a synthetic dataset for a three-class classification problem. This data generation approach ensures a certain degree of separability between different classes while maintaining sufficient complexity to verify the robustness of the TTA classifier under noisy environments. The 2-dimensional data visualization is shown in Fig.~\ref{fig:synthetic}(a).

In the experimental design, we randomly generate 4-dimensional data and conduct experiments using 4 qubits. Each data point is encoded into a quantum state through $R_y$ encoding gates. The dataset contains 2000 training samples and 1000 test samples per class, resulting in a total of 6000 training samples and 3000 test samples. Based on the principle of maximizing trace distance between positive and negative class average states, we construct a trace distance binary tree, whose structure is shown in Fig.~\ref{fig:synthetic}(b), with the trace distance values between average quantum states of different classes presented in Table~\ref{tab:synthetic_trace_distances}.

The training procedure for the TTA classifier follows the same protocol as described in Subsec.~\ref{subsec:mnist_dataset} and Subsec.~\ref{subsec:annni_model}. Specifically, we use the PQC architecture shown in Fig.~\ref{fig:arch} with circuit depth $L=20$, the Pauli Z operator on the first qubit as the observable, hinge loss as the loss function, a batch size of 200, the Adam optimizer~\cite{kingma2014adam} with a learning rate of 0.005, and the same early stopping strategy.

\begin{table}[h]
    \centering
    \caption{Trace distances between the average quantum states of the negative class (left) and positive class (right) for each node in the synthetic classification dataset TTA tree.}
    \label{tab:synthetic_trace_distances}
    \begin{tabular}{cccc}
    \hline
    \hline
    Node & Classes  & Trace Distance & Training Samples \\
    \hline
    Node 1 & \{2\} vs \{0,1\} & 0.6700 & 6000 \\
    Node 2 & \{0\} vs \{1\} & 0.5913 & 4000 \\
    \hline
    \hline
    \end{tabular}
\end{table}

\subsection{Performance under Noise}\label{subsec:performance_noisy}

In quantum computing, noise is inevitable and affects quantum state evolution and measurement results. We consider three typical quantum noise models.

\subsubsection{Noise Models}

\textbf{Generalized Amplitude Damping Noise.} Generalized Amplitude Damping noise~\cite{nielsen2000quantum} describes the interaction between quantum systems and finite-temperature environments, considering both energy dissipation and thermal excitation processes. This noise is described by four Kraus operators:
$$
\begin{aligned}
    E_0 =\sqrt{p}\begin{pmatrix} 0 & \sqrt{\gamma} \\ 0 & 0 \end{pmatrix}, E_2 = \sqrt{1-p}\begin{pmatrix} \sqrt{1-\gamma} & 0 \\ 0 & 1 \end{pmatrix} \\
    \quad E_1 =  \sqrt{p}\begin{pmatrix} 1 & 0 \\ 0 & \sqrt{1-\gamma} \end{pmatrix}  E_3 = \sqrt{1-p}\begin{pmatrix} 0 & 0 \\ \sqrt{\gamma} & 0 \end{pmatrix},
\end{aligned}
$$
where $\gamma \in [0,1]$ is the damping parameter and $p \in [0,1]$ is the excitation probability related to the environment temperature. When $p=1$, this noise reduces to standard amplitude damping noise.

\textbf{Depolarizing Noise.} Depolarizing noise~\cite{nielsen2000quantum} is one of the most common quantum noise models, describing the degradation of quantum states toward the maximally mixed state. For single-qubit systems, the Kraus operators of depolarizing noise are:
$$
\begin{aligned}
    &E_0 = \sqrt{1-\frac{3p}{4}}I, & E_1 = \sqrt{\frac{p}{4}}\sigma_x \\
    &E_2 = \sqrt{\frac{p}{4}}\sigma_y, & E_3 = \sqrt{\frac{p}{4}}\sigma_z
\end{aligned}
$$
where $p \in [0,1]$ is the noise strength parameter, $I$ is the identity matrix, and $\sigma_x, \sigma_y, \sigma_z$ are Pauli matrices. This noise randomly applies Pauli operators to the quantum state with probability $p$ and leaves the quantum state unchanged with probability $1-p$.

\textbf{Reset Error Noise.} Reset Error noise~\cite{nielsen2000quantum} describes the process where qubits are accidentally reset to a fixed state, which frequently occurs in actual quantum devices. The Kraus operators for this noise are:
$$
\begin{aligned}
    &E_0 = \sqrt{1-p}I, \\ 
    & E_1 = \sqrt{p}|0\rangle\langle 0|, \quad E_2 = \sqrt{p}|0\rangle\langle 1|,
\end{aligned}
$$
where $p \in [0,1]$ is the reset probability. This noise forcibly resets the qubit to the $|0\rangle$ state with probability $p$ and leaves the quantum state unchanged with probability $1-p$.

In our experiments, we set the noise strength parameter to $p=0.1$ for depolarizing noise. For generalized amplitude damping noise, we set $\gamma=0.05$ and $p=0.05$. For reset error noise, we set $p=0.05$. These parameter values reflect typical noise levels in actual quantum devices.

\subsubsection{Results and Analysis}

\begin{figure}[H]
    \centering
    \includegraphics[width=0.48\textwidth]{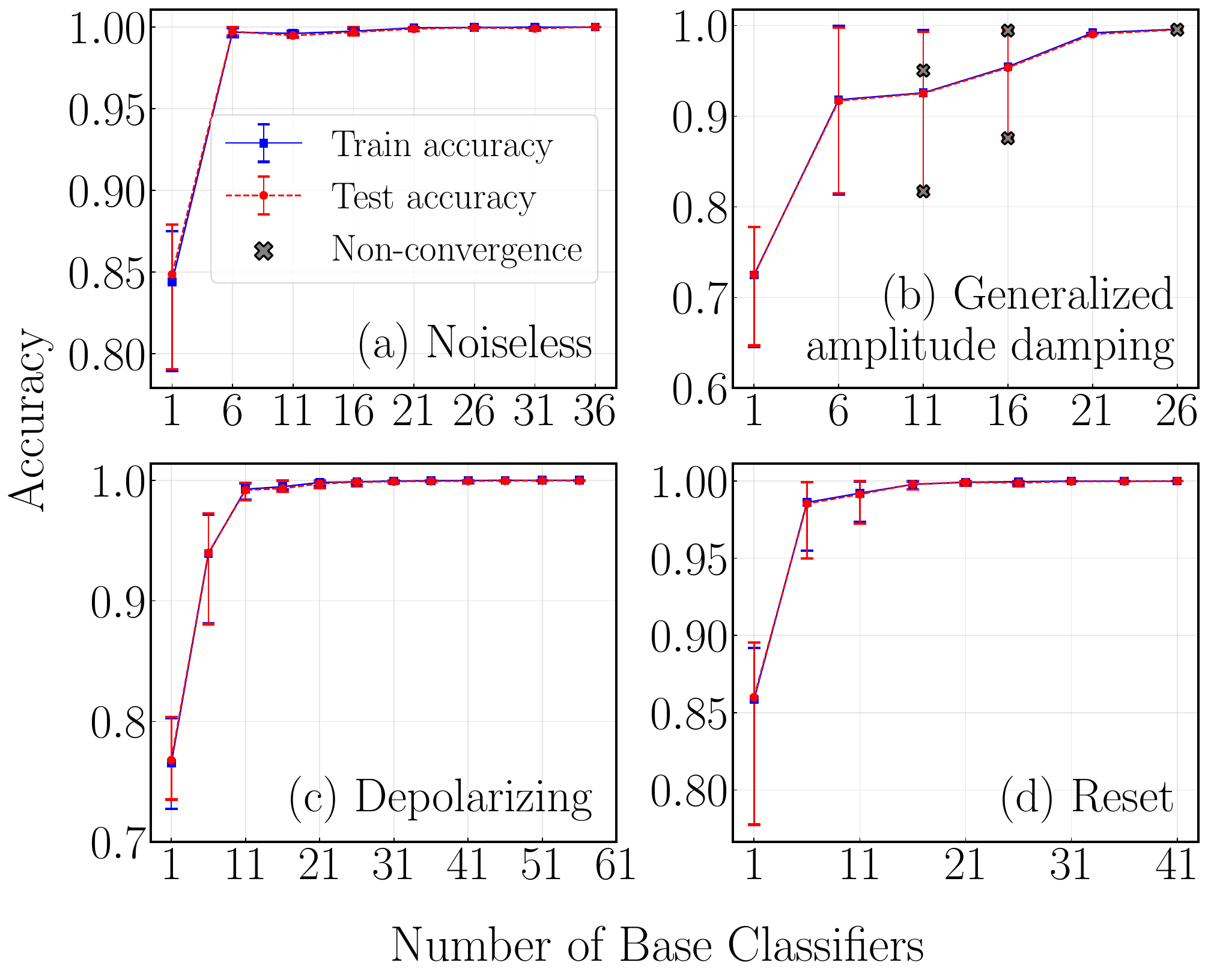}
    \caption{Classification results of TTA classifier on synthetic classification data under different conditions: (a) Noiseless case (b) Generalized amplitude damping noise case, where gray points indicate that some member classifiers failed to converge and exited training directly due to their base classifiers having train errors greater than 0.5. (c) Depolarizing noise case (d) Reset error case.  Lines show average of 5 independent runs with error bars indicating min/max values, gray points represent the final results of classifiers that failed to converge.}
    \label{fig:noise}
\end{figure}

Fig.~\ref{fig:noise} demonstrates the classification performance of the TTA classifier under different noise conditions. Under noiseless conditions (Fig.~\ref{fig:noise}(a)), the TTA classifier converges stably with both training and test accuracies reaching 1.0, demonstrating the ideal performance of the algorithm. Under generalized amplitude damping noise (Fig.~\ref{fig:noise}(b)), some member classifiers fail to converge due to base classifier train errors exceeding 0.5 (indicated by gray points in the figure), leading to some degradation in overall performance, though it still maintains a relatively high accuracy. In contrast, depolarizing noise (Fig.~\ref{fig:noise}(c)) and reset error noise (Fig.~\ref{fig:noise}(d)), while slowing down convergence speed and increasing the required number of base classifiers, can still achieve classification accuracies close to 1. This phenomenon is consistent with the theoretical expectations of the AdaBoost algorithm: as long as the base classifier performance is slightly better than random guessing, the algorithm can guarantee convergence, with noise primarily affecting convergence rate rather than final performance. Notably, even under the generalized amplitude damping noise conditions where convergence fails (Fig.~\ref{fig:noise}(b)), the TTA classifier still demonstrates good robustness, providing important practical support for applications on actual quantum devices.

\subsubsection{Impact of Circuit Depth on Amplitude Damping Noise}

\begin{figure}[H]
    \centering
    \includegraphics[width=0.49\textwidth]{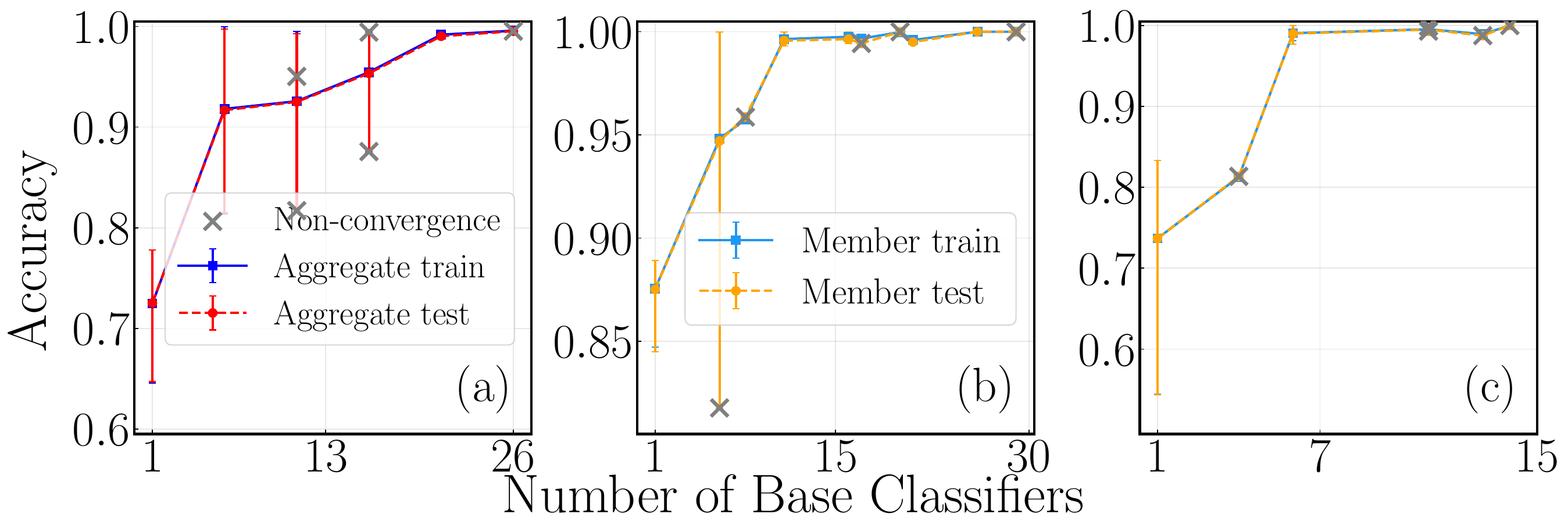}
    \caption{Under generalized amplitude damping noise conditions with circuit layer $L = 20$, (a) although not all 5 trials fully converged to perfect accuracy, some still achieved training and test accuracy close to 1. (b)-(c) Individual member classifiers exhibit similar behavior. 
    Lines show average of 5 independent runs with error bars indicating min/max values, gray points represent the final results of classifiers that failed to converge}
    \label{fig:amp}
\end{figure}

\begin{figure}
    \centering
    \includegraphics[width=0.49\textwidth]{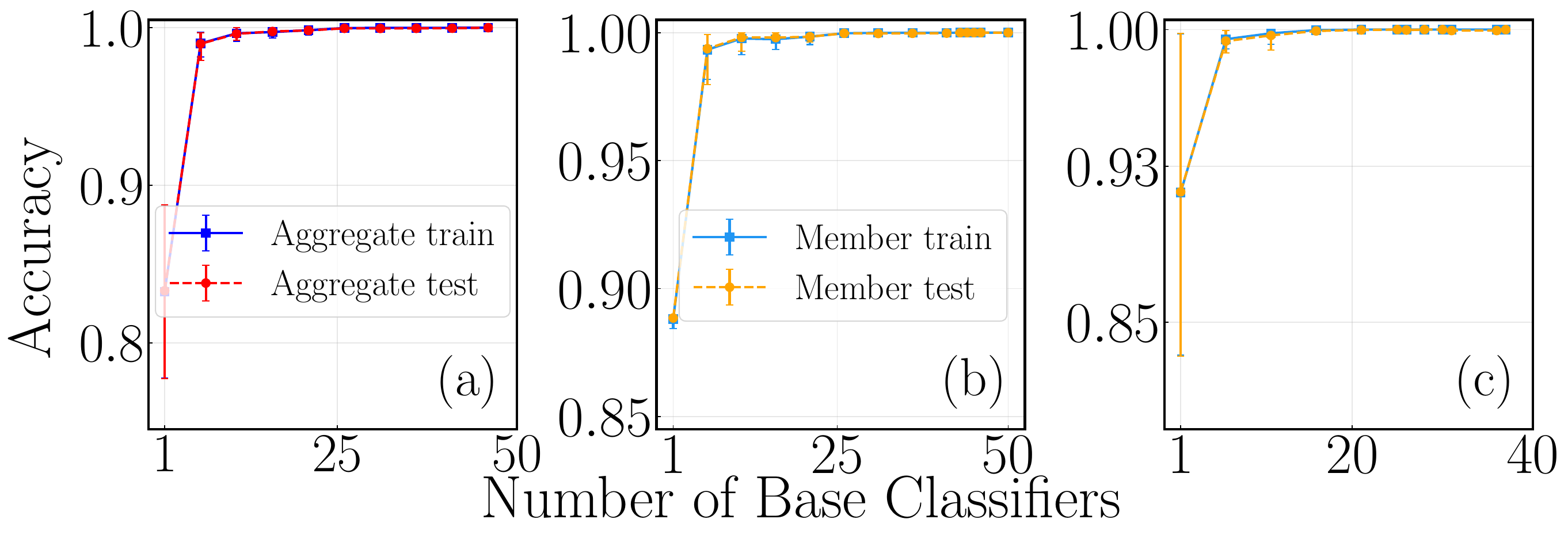}
    \caption{Under generalized amplitude damping noise conditions with circuit layer $L = 10$, (a) as the number of base classifiers increases, both training and test accuracy approach 1. (b)-(c)  Similar trend for individual member classifiers. Lines show average of 5 independent runs with error bars indicating min/max values, gray points represent the final results of classifiers that failed to converge}
    \label{fig:amp_10}
\end{figure}

As shown in Fig.~\ref{fig:noise}(b), under generalized amplitude damping noise, the TTA classifier failed to converge in all five independent runs. A closer examination (Fig.~\ref{fig:amp}) reveals that both member classifiers in trace distance tree (as shown in Fig.~\ref{fig:synthetic}(b)) failed to converge in each run: while some achieved accuracies approaching 0.99, others did not even reach 0.9. To mitigate this issue, we reduced the quantum circuit depth from $L=20$ to $L=10$. As shown in Fig.~\ref{fig:amp_10}, with the reduced circuit depth, all member classifiers across every experiment achieved 100\% train and test accuracy as the number of base classifiers increased, and the TTA classifier exhibited perfect convergence. This improvement stems from two factors: first, the shallower circuit significantly reduces noise accumulation; second, even with only 10 layers, the TTA classifiers still function as weak classifiers (i.e., train error below 0.5), thereby preserving the theoretical convergence guarantees of AdaBoost.

\subsection{Comparison with OVO and OVR}\label{subsec:comparison_ovo_ovr}

\subsubsection{Comparison of Classifier Complexity}

\begin{figure}[H]
    \centering
    \includegraphics[width=0.45\textwidth]{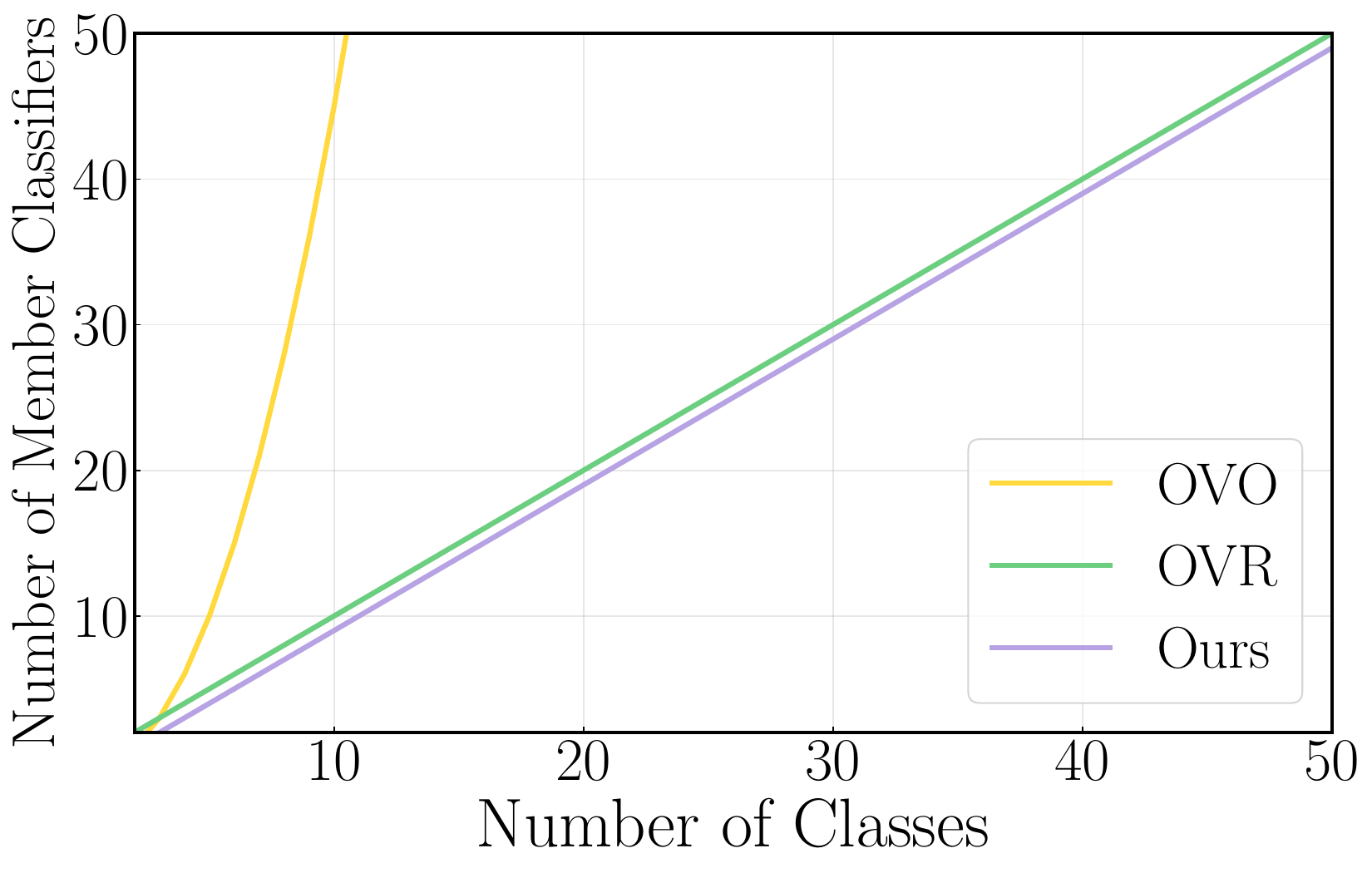}
    \caption{Comparison of the number of member classifiers required as the number of classes $K$ increases for the TTA classifier, OVR, and OVO classifiers.}
    \label{fig:K_comparison}
\end{figure}
To demonstrate the effectiveness of the TTA classifier and its advantages in training resource consumption, we compare it with the mainstream OVO (One-vs-One) and OVR (One-vs-Rest) methods. The TTA classifier is organized as a full binary tree, where each internal node has two child nodes representing the positive and negative classes, respectively. For a tree with $K$ leaf nodes (corresponding to $K$ classes), there are exactly $K-1$ internal nodes, requiring only $K-1$ member classifiers. In contrast, the OVR method constructs a separate classifier for each class (positive vs.\ all others as negative), requiring $K$ member classifiers in total. The OVO method trains one classifier for every pair of classes, resulting in $K(K-1)/2$ member classifiers. As shown in Fig.~\ref{fig:K_comparison}, the number of member classifiers required by each method varies with the number of classes. Notably, as the number of classes increases, the number of classifiers required by OVO grows rapidly, while the TTA classifier consistently requires one fewer member classifier than OVR.

\subsubsection{Trace Distance Analysis of Class Partitions}

Tables~\ref{tab:OVR} and~\ref{tab:OVO} present the class partitions and corresponding  trace distances between positive and negative class average states for the OVR and OVO classifiers. The TTA tree partitioning ($\{2\} \text{vs.} \{0,1\}$ followed by $\{0\} \text{vs.} \{1\}$) achieves the largest trace distance among all possible binary tree configurations. Consider the alternative where the first node is $\{1\} \text{vs.} \{0,2\}$: although the subsequent split between $\{0\}$ and $\{2\}$ yields a trace distance of 0.8379, the initial split has a trace distance of only 0.4406 with 6000 training samples, making convergence considerably more difficult. Similarly, starting with $\{0\}  \text{vs.} \{1,2\}$ and then separating classes 1 and 2 results in lower trace distances for both classifiers compared to the TTA partitions. Thus, the TTA partitioning strategy effectively balances classification difficulty and sample distribution, leading to more efficient training.

\begin{table}[h]
    \centering
    \caption{Trace distances between the average quantum states of the negative class (left) and positive class (right) for each member classifier in the OVR reduction method.}
    \label{tab:OVR}
    \begin{tabular}{ccc}
    \hline
    \hline
    Classifier & Trace Distance & Training Samples \\
    \hline
    Class 0 vs Rest & 0.6498 & 6000 \\
    Class 1 vs Rest & 0.4406 & 6000 \\
    Class 2 vs Rest & 0.6700 & 6000 \\
    \hline
    \hline
    \end{tabular}
\end{table}

\begin{table}[h]
    \centering
    \caption{Trace distances between the average quantum states of the negative class (left) and positive class (right) for each member classifier in the OVO reduction method.}
    \label{tab:OVO}
    \begin{tabular}{ccc}
    \hline
    \hline
    Classifier & Trace Distance & Training Samples \\
    \hline
    Class 0 vs Class 1 & 0.5913 & 4000 \\
    Class 0 vs Class 2 & 0.8379 & 4000 \\
    Class 1 vs Class 2 & 0.5739 & 4000 \\
    \hline
    \hline
    \end{tabular}
\end{table}

\subsubsection{Performance Comparison}

\begin{figure}
    \centering
    \includegraphics[width=0.49\textwidth]{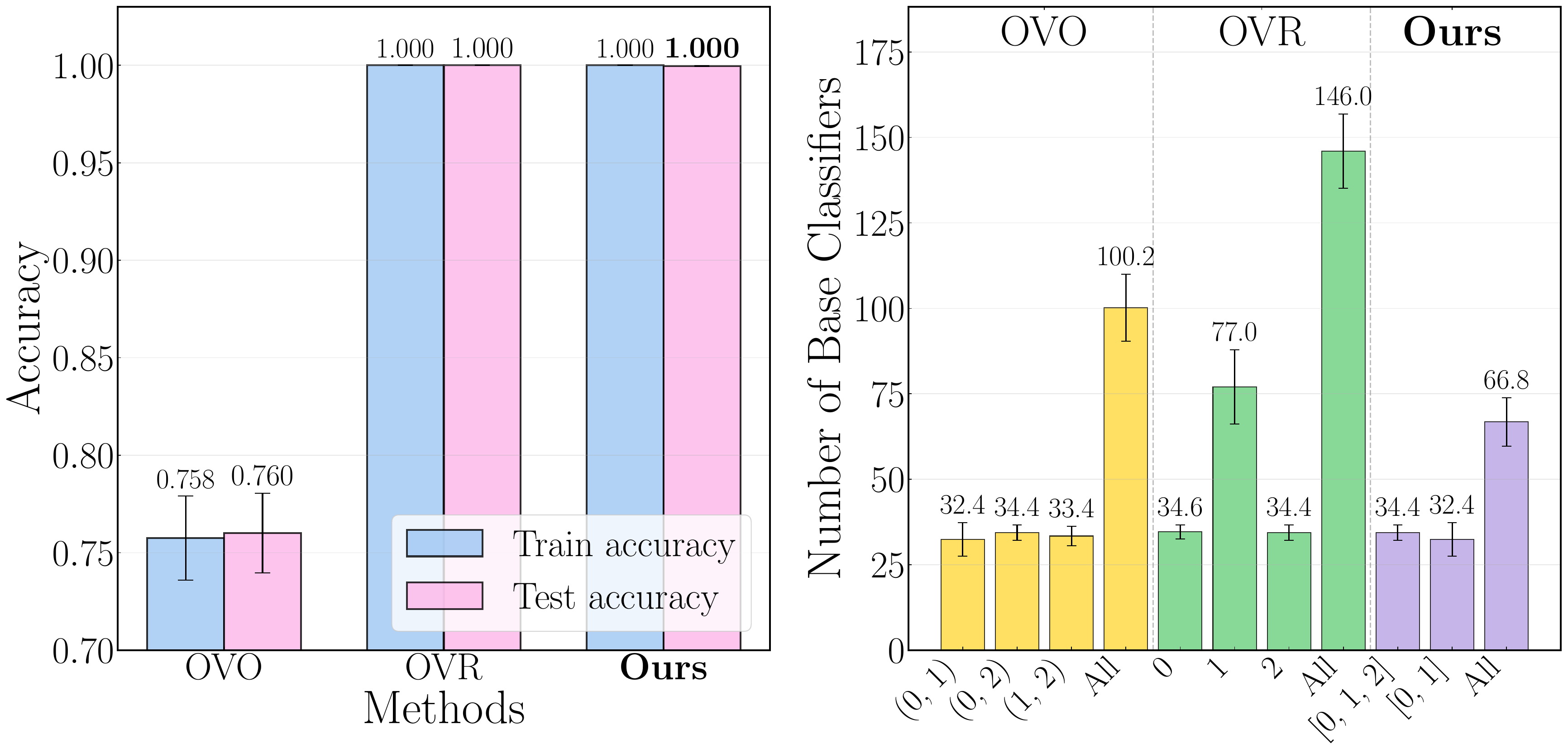}
    \caption{Comparison of train and test accuracy, as well as the number of base classifiers used, between the TTA classifier and the OVO and OVR methods: (a) Train and test accuracy comparison, (b) Number of base classifiers comparison. The numbers on the bars represent the average over 5 independent runs, and the error bars indicate the maximum and minimum values across these runs.}
    \label{fig:OVO_OVR_comparison}
\end{figure}

As shown in Fig.~\ref{fig:OVO_OVR_comparison}, the OVO method performs significantly worse than the TTA classifier. This is because OVO trains a separate classifier for each pair of classes, resulting in completely independent binary classifiers that lack global structural information across the dataset. Consequently, final predictions rely on  maximum confidence aggregation, which increases inference uncertainty and reduces robustness to noise, leading to lower accuracy and stability.

In contrast, the OVR method achieves classification accuracy comparable to ours. However, a closer examination of base classifier efficiency reveals significant resource waste. For instance, when class 1 serves as the positive class against $\{0,2\}$ as the negative, the trace distance between their average states is relatively small. As a result, the corresponding binary AdaBoost classifier requires substantially more base classifiers to achieve the same train error convergence bound.

Compared with these two alternatives, our trace distance tree partitioning strategy offers significant theoretical and practical advantages. By recursively partitioning the multi-class problem using a binary tree structure, each internal node groups classes to maximize the average quantum state trace distance, thereby ensuring better class separability while keeping the difficulty of each binary task manageable. Furthermore, this hierarchical structure naturally captures global class relationships, enabling efficient information transfer and decision without the need to construct redundant base classifiers for every class or pair of classes. As demonstrated by our experimental results, the TTA classifier substantially reduces both the number of base classifiers and the training resources required, all without sacrificing classification accuracy. This combination of efficiency and scalability makes it particularly well-suited for practical deployment in scenarios where quantum computing resources are limited.

\section{Encoding Matters}\label{sec:encoding}

\begin{table*}[ht]
    \centering
    \caption{Impact of trace distance between average states of two classes in the training set on AdaBoost results. }
    \label{tab:encoding_trace_distance}
    \begin{tabular}{cccccc}
        \hline
        \hline
        Dataset & Encoding & Dimension & Average & Ensemble Accuracy & Termination \\
        (Classes) & Method & (Qubits) & Trace Distance & (Train/Test) & Reason\\
        \hline
        CIFAR10 (0-2) & Amplitude & 256 (8) & \cellcolor{red!20}0.0760 (2000) & \cellcolor{red!20}72.58 $\pm$ 5.63\% / 67.52 $\pm$ 3.91\% & {\color{orange}$\blacksquare$} {\color{red}$\blacksquare$} {\color{orange}$\blacksquare$} {\color{orange}$\blacksquare$} {\color{red}$\blacksquare$} \\
        FashionMNIST (1-9) & Amplitude & 256 (8) & \cellcolor{green!20}0.8343 (2000) & \cellcolor{green!20}100.00$\pm$0.01\% / 99.91$\pm$0.06\% & {\color{green}$\blacksquare$} {\color{green}$\blacksquare$} {\color{green}$\blacksquare$} {\color{green}$\blacksquare$} {\color{green}$\blacksquare$} \\
        Two curves & Amplitude & 10 (4) & \cellcolor{orange!20}0.3856 (120) & \cellcolor{green!20}99.83$\pm$0.33\% / 100.00$\pm$0.00\% & {\color{green}$\blacksquare$} {\color{green}$\blacksquare$} {\color{green}$\blacksquare$} {\color{red}$\blacksquare$} {\color{red}$\blacksquare$} \\
        Hidden Manifolds & Amplitude & 2 (1) & \cellcolor{red!20}0.1285 (120) & \cellcolor{red!20}70.00$\pm$3.16\% / 41.67$\pm$1.83\% & {\color{red}$\blacksquare$} {\color{orange}$\blacksquare$} {\color{orange}$\blacksquare$} {\color{red}$\blacksquare$} {\color{red}$\blacksquare$} \\
        Hidden Manifolds & Angle & 2 (1) & \cellcolor{red!20}0.2344 (120) & \cellcolor{red!20}79.08$\pm$7.06\% / 67.33$\pm$1.70\% & {\color{red}$\blacksquare$} {\color{orange}$\blacksquare$} {\color{red}$\blacksquare$} {\color{red}$\blacksquare$} {\color{red}$\blacksquare$} \\
        Linearly Separable & Amplitude & 2 (1) & \cellcolor{red!20}0.0746 (120) & \cellcolor{red!20}71.58$\pm$5.09\% / 42.33$\pm$3.27\% & {\color{red}$\blacksquare$} {\color{orange}$\blacksquare$} {\color{red}$\blacksquare$} {\color{red}$\blacksquare$} {\color{orange}$\blacksquare$} \\
        Linearly Separable & Angle & 2 (1) & \cellcolor{orange!20}0.4023 (120) & \cellcolor{green!20}100.00$\pm$0.00\% / 99.67$\pm$0.67\% & {\color{green}$\blacksquare$} {\color{green}$\blacksquare$} {\color{red}$\blacksquare$} {\color{green}$\blacksquare$} {\color{green}$\blacksquare$} \\
        \hline
        \hline
    \end{tabular}
\end{table*}

In the previous sections, we showed that the TTA classifier can achieve 100\% train accuracy and nearly 100\% test accuracy on multi-class tasks, for both classical and quantum data. This strong performance comes from using shallow base classifier circuits to avoid barren plateaus, and combining multiple shallow base classifiers to approximate the global optimum. As a result, TTA overcomes barren plateaus and spurious local minima in training while maintaining good generalization. However, a fundamental limitation remains: how to encode classical data into quantum states. Previous studies have shown that commonly used encoding methods, including angle encoding~\cite{li2022concentration} and amplitude encoding~\cite{wang2025limitations}, have inherent limitations that can lead to poor performance in QML models.
Furthermore, Ref.~\cite{wang2025limitations} shows that as the trace distance between the average states of two classes decreases, training becomes increasingly difficult and predictions tend toward random guessing.

To investigate the effect of encoding on TTA training and test, we selected the following binary datasets, where "Two curves", "Hidden Manifolds", and "Linearly Separable" are directly provided by PennyLane~\cite{bowles2024better}:
\begin{itemize}
    \item CIFAR10~\cite{krizhevsky2009learning}: A widely-used benchmark dataset for image classification, consisting of 60,000 32$\times$32 color images across 10 classes (airplane, automobile, bird, cat, deer, dog, frog, horse, ship, and truck). We select classes 0 (airplane) and 2 (bird) for binary classification.
    \item FashionMNIST~\cite{xiao2017fashion}: A dataset of Zalando's article images, containing 70,000 28$\times$28 grayscale images across 10 fashion categories (T-shirt/top, trouser, pullover, dress, coat, sandal, shirt, sneaker, bag, and ankle boot). We select classes 1 (trouser) and 9 (ankle boot) for binary classification. 
    \item Two curves: This data is from the benchmark~\cite{bowles2024better,buchanan2020deep}, where two 1-dimensional class manifolds are embedded into a \(d=10\)-dimensional space using a low-degree Fourier series of a fixed degree \(D=5\). The two classes are separated by a constant offset \(\Delta = 0.1\), with additive Gaussian noise, creating a controlled benchmark for probing neural network performance on complex data geometries.
    \item Hidden Manifolds: This data is generated using the hidden manifold procedure by~\cite{goldt2020modeling} from benchmark~\cite{buchanan2020deep}, where inputs are created on a low-dimensional manifold, labeled by a simple randomly-initialized neural network, and then projected into a high-dimensional ambient space. We specifically utilize the benchmark from this framework where the ambient dimension is fixed at \(d=2\) and the manifold dimension is fixed at \(m=6\).
    \item Linearly Separable: This data is generated from the canonical ``fruit-fly'' example of a linearly separable problem~\cite{rosenblatt1958perceptron}, taken from the benchmark~\cite{buchanan2020deep}. Inputs are sampled uniformly from a 2-dimensional hypercube and classified by a hyperplane orthogonal to the vector $\boldsymbol{w} = [1,1]^T$. A data-free margin $\Delta = 0.02$ around the hyperplane ensures that all datapoints $\boldsymbol{x}$ satisfy $|\boldsymbol{w}^{\top} \boldsymbol{x}| > \Delta$.
\end{itemize}

Table~\ref{tab:encoding_trace_distance} summarizes the binarydataset (with selected class labels in parentheses), encoding method (angle encoding is implemented by mapping data to single-qubit gates $R_y, R_x$), data dimension (with the number of qubits in parentheses), trace distance between average quantum states of the two classes (with the number of training samples per class in parentheses), final AdaBoost ensemble train and test accuracies (mean $\pm$ standard deviation), and the stopping criterion for each base classifier: {\color{red}$\blacksquare$} indicates train error exceeded 0.5; {\color{orange}$\blacksquare$} indicates the maximum number of base classifiers (500) was reached; and {\color{green}$\blacksquare$} means improvement in train error fell below 0.005.

As shown in Table~\ref{tab:encoding_trace_distance}, when the average trace distance is large (e.g., FashionMNIST), AdaBoost achieves perfect fitting even with 4000 training samples (2000 per class). For datasets with moderate trace distance, convergence behavior varies across five independent runs. Sometimes the train error lower bound drops below 0.005, while other times the final base classifier's train error remains above 0.5. When the trace distance is small, the ensemble typically reaches the maximum number of base classifiers (500) or fails to converge entirely. Remarkably, even for datasets with moderate trace distance and only 240 training samples (120 per class), achieving 100\% train accuracy still yields test accuracy exceeding 99\%, highlighting the strong generalization capability of QML methods.

It is worth noting that some studies have suggested that the limitations caused by amplitude encoding may be alleviated by angle encoding. Our results reveal a more nuanced picture. For the Linearly Separable dataset, switching from amplitude encoding to angle encoding increases the trace distance between average states from 0.0746 to 0.4023, enabling AdaBoost to achieve 100\% train accuracy and an average test accuracy of 99.67\%. In contrast, for the Hidden Manifolds dataset, angle encoding yields a trace distance of only 0.2344, which remains insufficient for AdaBoost to enhance trainability. These findings demonstrate that improving trainability by changing the encoding method is neither simple nor universally effective.

\section{The effect of Early Stopping}\label{sec:early_stopping}

In this section, we demonstrate the efficiency of the early stopping strategy. While the TTA classifier may appear to require training a large number of base classifiers (and hence many PQCs), early stopping ensures that each PQC typically converges in just over 10 epochs—far fewer than the hundreds of epochs often needed by conventional classifiers—thereby substantially reducing overall training costs.

Using the FashionMNIST dataset as an example (see Tab.~\ref{tab:encoding_trace_distance}), we fix the initialization parameters of the first base classifier and compare training with and without early stopping. All other hyperparameters—optimizer, batch size, and learning rate—remain unchanged. In each round, we select the parameters that achieve the lowest train error as the final parameters for the $t$-th base classifier. Consequently, even when starting from identical initial parameters, the optimal parameters selected in each round may differ, leading to varying train errors. These differences propagate through the ensemble: they affect the weight assigned to each base classifier and the subsequent reweighting of training samples, ultimately causing the overall performance to diverge even when all base classifiers share the same initialization.

\begin{figure}[htpb]
    \centering
    \includegraphics[width=0.48\textwidth]{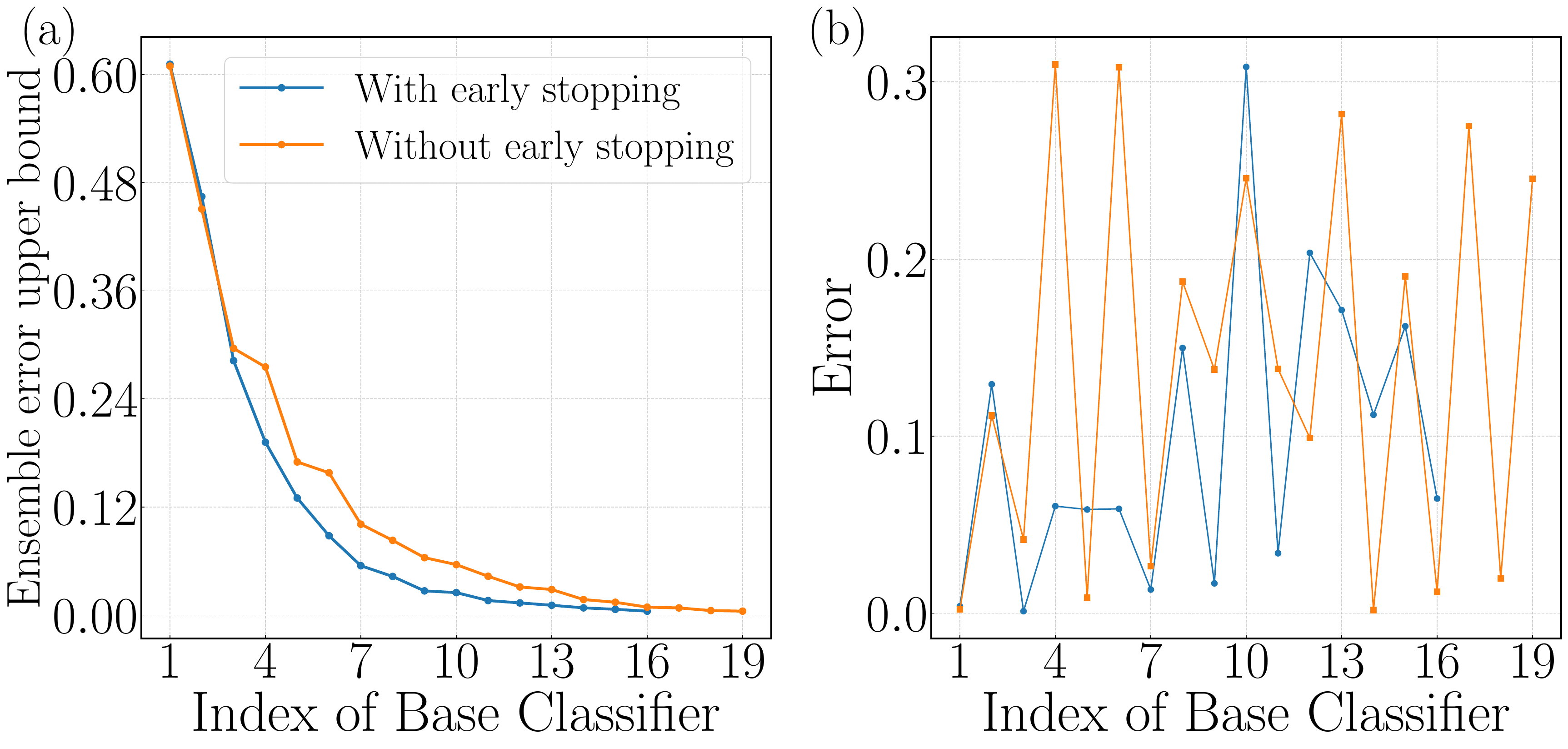}
    \caption{Comparison of (a) the ensemble train error upper bound and (b) the train error of the $t$-th base classifier, with and without early stopping.}
    \label{fig:early_stopping}
\end{figure}

We compared the ensemble train error upper bound at the $t$-th base classifier under two conditions: with and without early stopping. Intuitively, one might expect that allowing each base classifier more training epochs would yield lower train errors, enabling the ensemble to reach convergence thresholds (e.g., 0.005) more quickly. However, as shown in Fig.~\ref{fig:early_stopping}(a), the opposite is true: with early stopping, the ensemble train error upper bound drops below the convergence threshold sooner. Specifically, without early stopping, 19 base classifiers were required, whereas with early stopping, only 16 sufficed. A closer look at Fig.~\ref{fig:early_stopping}(b) reveals that although the train error of individual base classifiers is slightly lower without early stopping in the first two rounds, overtraining induces large fluctuations in the sample weight distribution in subsequent rounds. These fluctuations make it harder for new base classifiers to distinguish the reweighted data, increasing optimization difficulty and raising train errors. Consequently, even without early stopping, the quality of later base classifiers deteriorates, necessitating more classifiers to achieve ensemble convergence. Thus, appropriately applying early stopping not only reduces overall training resource consumption but also improves the ensemble's general performance.

\begin{figure}[htpb]
    \centering
    \includegraphics[width=0.36\textwidth]{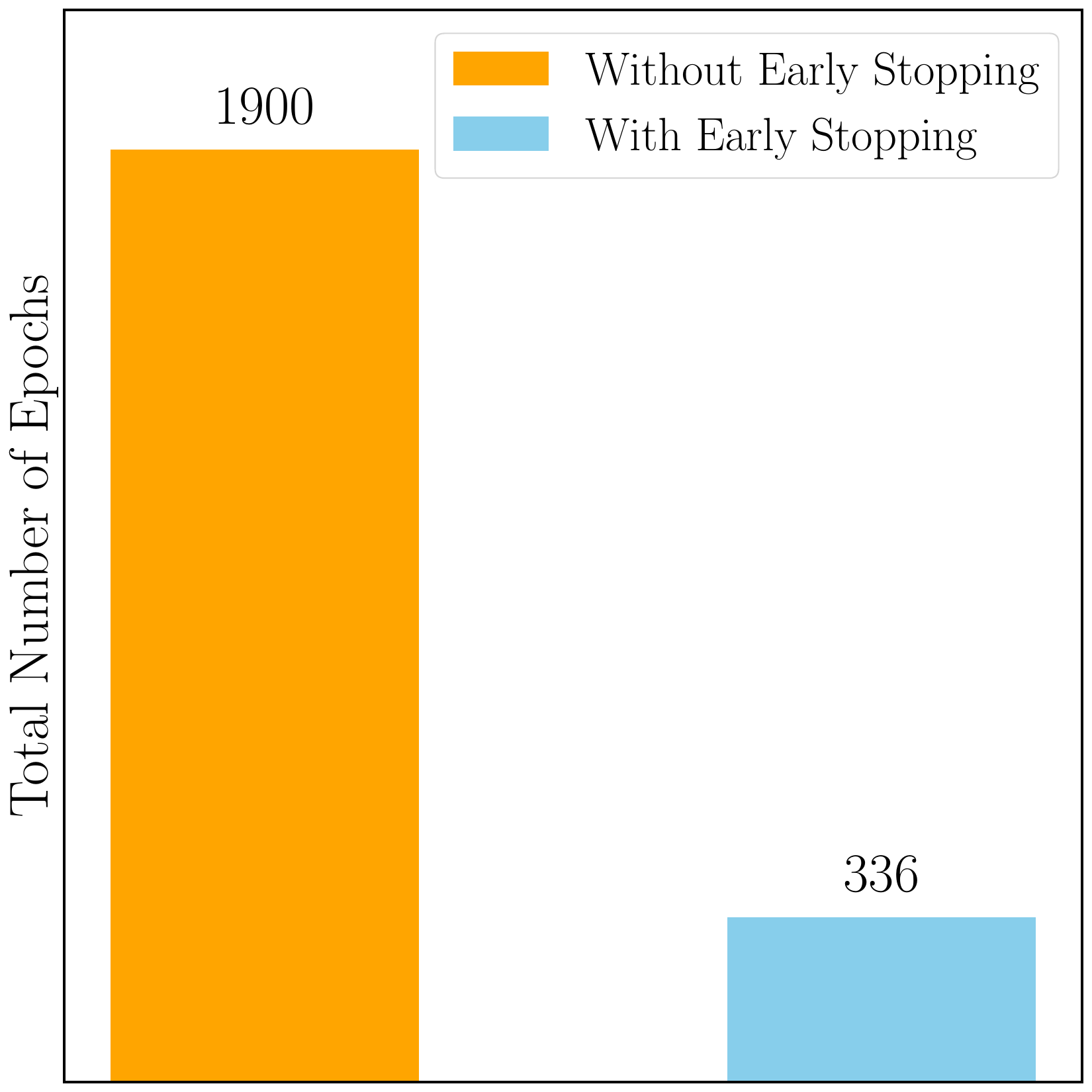}
    \caption{Total number of epochs required to reach the same ensemble train error upper bound, with and without early stopping.}
    \label{fig:early_stopping_epochs}
\end{figure}

Experimental results showed that both ensembles achieved 100\% accuracy on both the train and test sets, regardless of whether early stopping was applied. To more intuitively illustrate the resource savings afforded by early stopping, Fig.~\ref{fig:early_stopping_epochs} compares the total number of training epochs required under both settings. Without early stopping, 19 base classifiers consumed a total of 1900 epochs; with early stopping, only 336 epochs were needed—a reduction of approximately 5.65$\times$ in computational cost. Further details on the epoch-wise evolution of train error for all 16 (with early stopping) and 19 (without early stopping) base classifiers are provided in Appendix~\ref{sec:early_stopping_details}.

\section{Conclusion}\label{sec:conclusion}
We introduced TTA classifier, a compact and practical strategy that pairs a quantum-aware reduction rule (trace-distance maximization) with nodewise AdaBoost of shallow variational circuits to address critical trainability limits in multi-class QML. TTA achieves 100\% train accuracy and near-perfect test accuracy across classical and quantum benchmarks while substantially reducing monolithic circuit depth and parameter budgets; it also exhibits clear robustness under realistic quantum noise. Together with scalable heuristics (greedy split, early stopping, parallel node training) and extensive empirical validation, TTA offers a practical blueprint for deploying multi-class quantum classifiers on near-term devices and motivates further hardware demonstrations.

\section*{Code Availability}
The code for this paper is available at \url{https://github.com/SheffieldWang/Quantum_TTA}. The implementation is based on MindSpore Quantum~\cite{xu2024mindspore}.

\section*{Acknowledgements}
This work was sponsored by CPS-Yangtze Delta Region Industrial Innovation Center of Quantum and Information Technology-MindSpore Quantum Open Fund, and supported by NSFC (Grant No. 62173201) and the Innovation Program for Quantum Science and Technology (Grants No. 2021ZD0300200).

\bibliography{ref}

\onecolumngrid

\pagebreak

\appendix

\resetAppendixCounters{A}

\section{Illustration of Other Multi-Class to Binary Reduction Methods}
\label{app:reduction_methods}
\subsection{Bitwise Aggregate}
This method assigns each class a unique binary code and trains one classifier per bit position, thereby reducing multi-class classification to binary classification. For $K=4$, the class labels $\{0,1,2,3\}$ are encoded as two-bit binary numbers: 00, 01, 10, and 11 respectively. A separate binary classifier is then trained for each bit position. 

Consider the first-bit classifier: it predicts whether the leading bit of the encoding is 0 or 1. During training, samples from classes 0 and 1 (with encodings 00 and 01, both having leading bit 0) are grouped as the negative class, while samples from classes 2 and 3 (with encodings 10 and 11, both having leading bit 1) form the positive class. For the second-bit classifier, the grouping differs: classes 0 and 2 (encodings 00 and 10, trailing bit 0) become the negative class, and classes 1 and 3 (encodings 01 and 11, trailing bit 1) become the positive class. All classifiers are trained independently.

During inference, each classifier predicts its corresponding bit for a new sample. The predicted bits are concatenated to form a binary code, which is then decoded to obtain the final class label. For $K=4$, only $\lceil \log_2 4 \rceil = 2$ binary classifiers are needed. However, this approach does not account for quantum distinguishability (such as trace distance between average quantum states) between classes, and each classifier must process the entire dataset, which can make training difficult.

\begin{figure}[htpb]
    \centering
    \includegraphics[width=0.58\textwidth]{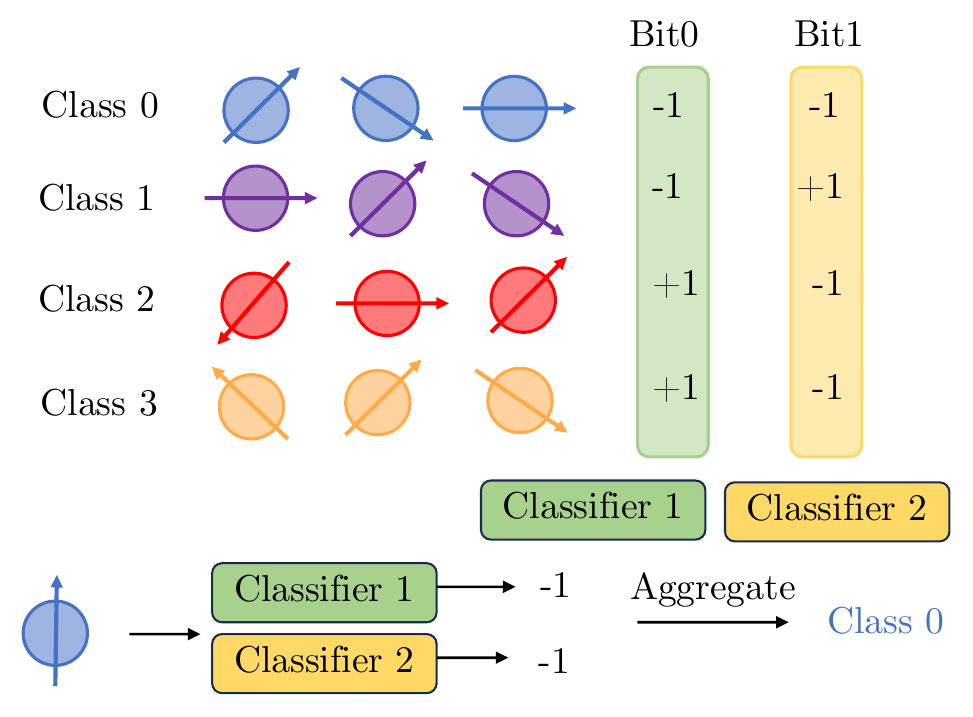}
    \caption{Illustration of the bitwise aggregate method for multi-class to binary reduction: left shows original multi-class labels, right shows their binary encoding. Each binary classifier predicts one bit, and the final label is decoded from the concatenated bits.}
    \label{afig:bitwise}
\end{figure}

\subsection{One-Versus-Rest (OVR)}

The One-Versus-Rest (OVR) approach is a widely used strategy for reducing multi-class classification to binary tasks. For a $K$-class problem, OVR trains $K$ binary classifiers, each distinguishing one class (labeled positive) from all others (labeled negative). At inference time, a new sample is passed through all $K$ classifiers, and the class whose classifier outputs the highest confidence is selected as the prediction.

Despite its simplicity, OVR has several drawbacks. First, each classifier faces a highly imbalanced dataset: the positive class comprises only $1/K$ of the samples, which can cause the model to overfit to the majority negative class and degrade discrimination. Second, the final multi-class prediction relies on comparing confidence from different binary classifiers, yet these scores often lack a unified standard and proper calibration, so directly comparing or aggregating them can introduce systematic bias. Third, OVR ignores quantum distinguishability measures such as trace distance between average quantum states, and every classifier must process the full dataset, increasing training difficulty.

\begin{figure}[htpb]
    \centering
    \includegraphics[width=0.6\textwidth]{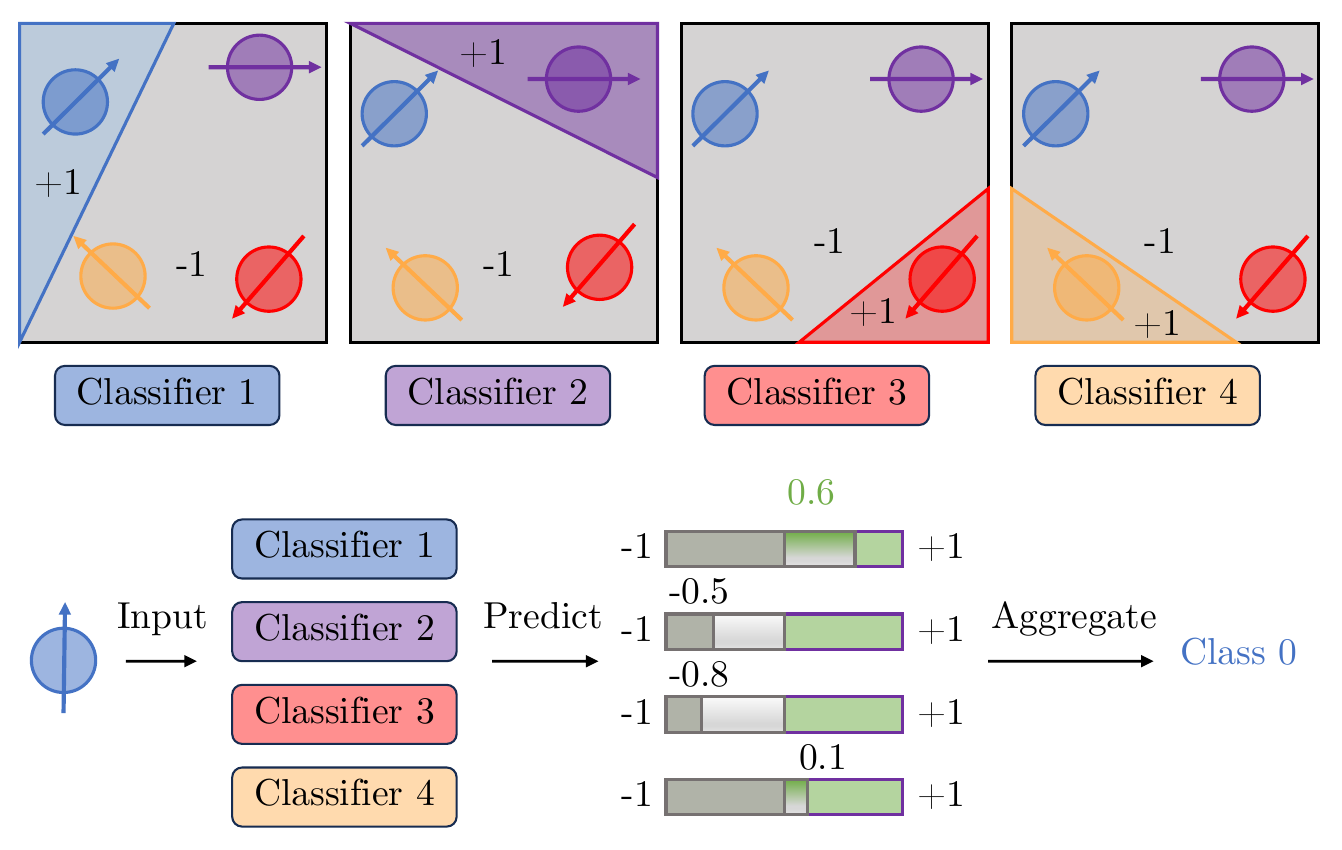}
    \caption{Illustration of the One-Versus-Rest (OVR) method: For each class, a binary classifier is trained to distinguish between that class and all other classes. During inference, a new sample is input to each member classifier, and the class with the highest confidence is selected as the predicted label.}
    \label{afig:ovr}
\end{figure}

\subsection{One-Versus-One (OVO)}
The One-Versus-One (OVO) method is another widely-used strategy for reducing multi-class classification to multiple binary problems. For a $K$-class problem, OVO trains $K(K-1)/2$ binary classifiers, one for each pair of distinct classes. Each classifier uses only data from its two designated classes, labeling one as positive and the other as negative. This design simplifies each classifier's task to distinguishing between just two classes, resulting in more balanced and cleaner training distributions. During inference, a new sample is passed through all classifiers, and the class with the highest confidence among all classifiers is selected as the final prediction.

However, OVO has notable drawbacks. As $K$ increases, the number of required classifiers grows quadratically, substantially raising computational costs for both training and inference. Moreover, since each classifier only sees data from two classes, it cannot leverage information from other classes. This limitation can lead to multiple classifiers assigning high confidence to the same sample, reducing the reliability of the final decision. Additionally, classifiers may behave unpredictably on samples from classes they have never encountered during training, further complicating prediction.
\begin{figure}[htpb]
    \centering
    \includegraphics[width=0.7\textwidth]{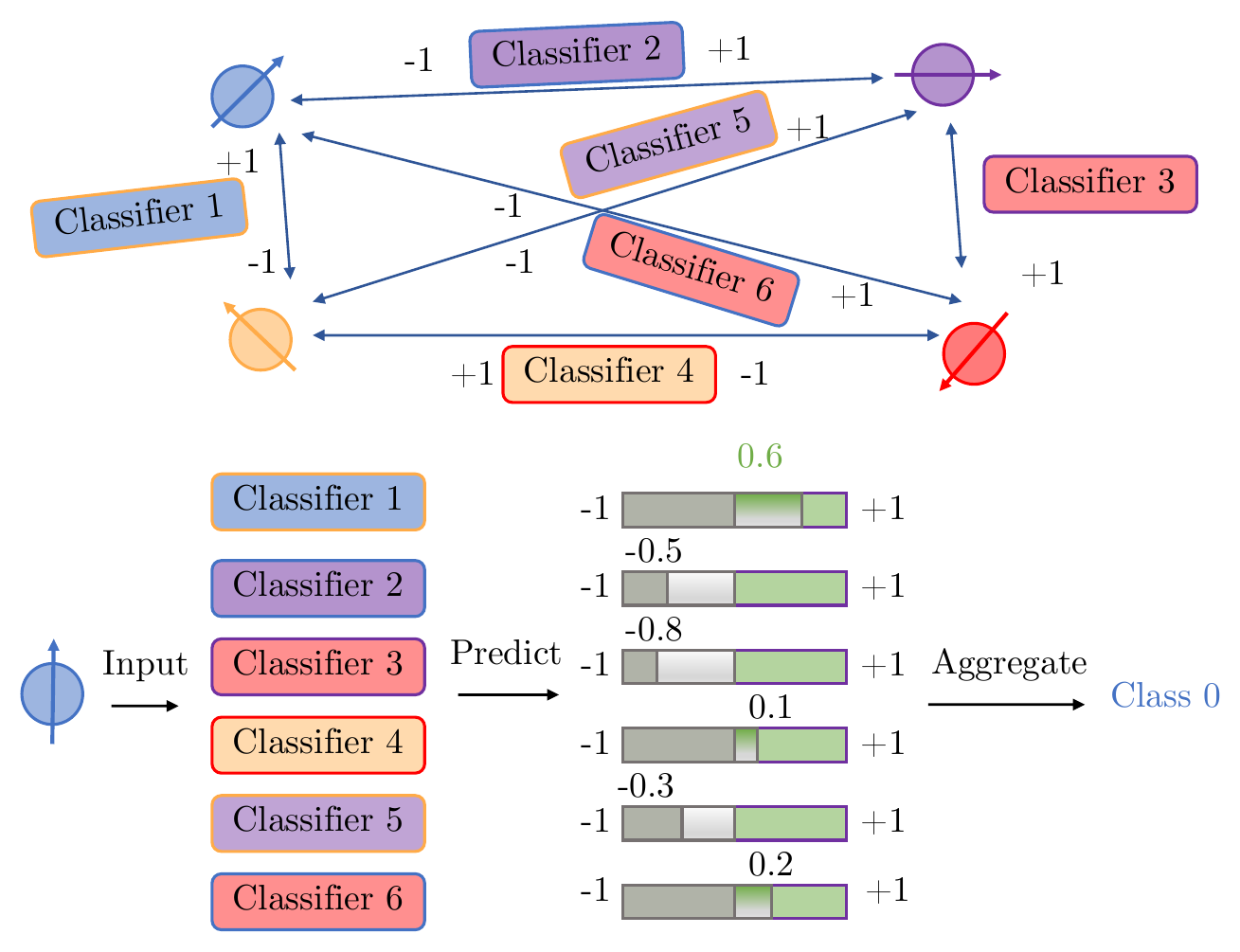}
    \caption{Illustration of the One-Versus-One (OVO) method: For each pair of classes, a binary classifier is trained to distinguish between them. During inference, a new sample is input to each member classifier, and the class with the highest confidence is selected as the predicted label.}
    \label{afig:ovo}
\end{figure}

\resetAppendixCounters{B}

\renewcommand{\thelemma}{A.\arabic{lemma}}
\renewcommand{\theproposition}{A.\arabic{proposition}}
\setcounter{theorem}{0}
\renewcommand{\thetheorem}{A.\arabic{theorem}}
\renewcommand{\thecorollary}{A.\arabic{corollary}}

\section{Other Algorithms}
\label{app:other_algorithms}
\subsection{Algorithm of Find Maximally Separated Binary Partition}

This algorithm, called \texttt{MaxBinarySplit}, partitions the current class set $K_{\text{cur}}$ into two subsets $(K_1, K_2)$ such that the average trace distance between them is maximized. By leveraging the trace distance between the average quantum states $\overline{\rho}_k$ of each class, this approach ensures that the resulting groups are as distinguishable as possible.

The procedure is as follows: First, for every pair of classes $(i, j)$ in the current class set $K_{\text{cur}}$, compute the trace distance between their average quantum states, given by $D_{ij} = \frac{1}{2}\|\overline{\rho}_i - \overline{\rho}_j\|_{1}$. Next, identify the pair with the largest trace distance, denoted as $(i^*, j^*)$, and use them to initialize the two subsets: $K_1 = \{i^*\}$ and $K_2 = \{j^*\}$. Then, for each remaining class $k \in K_{\text{cur}} \setminus \{i^*, j^*\}$, assign it to the subset whose initial element ($i^*$ or $j^*$) has the smaller trace distance to $k$. Finally, output the resulting partition $(K_1, K_2)$, which greedily attempts to find a partition such that the average quantum states of the two groups have a large trace distance.

\begin{algorithm}[H]
    \caption{MaxBinarySplit}\label{alg:max_sep_partition}
    \SetKwInOut{Input}{Input}
    \SetKwInOut{Output}{Output}
    \SetKwFunction{MaxBinarySplit}{MaxBinarySplit}
    \SetKwComment{tcp}{// }{}
    
    \Input{Set of average quantum states for each class in the current subset $\left\{\overline{\rho}_k\right\}_{k \in K_{\text{cur}}}$, class label set $K_{\text{cur}}$}
    \Output{A binary partition $(K_1, K_2)$ that maximizes the average trace distance between the two groups}
    
    \BlankLine
    Compute the trace distance between all pairs of classes:\\
    \ForEach{classes $i, j \in \mathcal{K},\, i < j$}{
        $D_{ij} \gets \frac{1}{2}\|\overline{\rho}_i - \overline{\rho}_j\|_{1}$\;
    }
    Select the class pair with the largest $D_{ij}$: $(i^*, j^*) = \arg\max_{i \neq j} D_{ij}$\;
    $K_1 \gets \{i^*\}$\\
    $K_2 \gets \{j^*\}$\\
    \ForEach{class $k \in \mathcal{K}$, $k \neq i^*, j^*$}{
        \uIf{$D_{k,i^*} < D_{k,j^*}$}{
            $K_1 \gets K_1 \cup \{k\}$\;
        }
        \Else{
            $K_2 \gets K_2 \cup \{k\}$\;
        }
    }
    \Return $(K_1, K_2)$
\end{algorithm}

\subsection{Multi-Class AdaBoost Algorithm}

Multi-class AdaBoost extends the original ensemble method to $K$-class tasks. Like its binary counterpart, it iteratively trains weak classifiers and adjusts sample weights to progressively enhance overall performance. However, this extension no longer retains the theoretical guarantee of an exponentially decreasing train error bound. The complete algorithm is outlined in Algo.~\ref{algo:multi_class_adaboost}.

\begin{algorithm}[H]
    \caption{Multi-Class Quantum AdaBoost}
    \label{algo:multi_class_adaboost}
    \SetKwInOut{Input}{Input}
    \SetKwInOut{Output}{Output}
    
    \Input{
        Training set $S = \left\{ \left( \rho^{(m)}, y^{(m)} \right) \right\}_{m=1}^M$ with $K$ classes;\\
        Maximum boosting rounds $T$
    }
    \Output{Final hypothesis $H_S$}
    
    Initialize weights for all samples: $\boldsymbol{w}_1(m) \gets \frac{1}{M}$, for $m \in [M]$\;
    \For{$t = 1$ \KwTo $T$}{
        Train base classifier $h_S^{(t)}$ on $S$ using weights $\boldsymbol{w}_t$\;
        Compute error: $\epsilon_t = \sum_{m=1}^M \boldsymbol{w}_t(m) \mathbbm{1}\left(y^{(m)} \neq h_S^{(t)}(\rho^{(m)})\right)$\;
        \If{$\epsilon_t \geq \frac{K-1}{K}$}{
            \textbf{break}
        }
        Calculate coefficient of base classifier $h_S^{(t)}$: $\alpha_t \gets \log\frac{1-\epsilon_t}{\epsilon_t} + \log(K-1)$\;
        \For{$m = 1$ \KwTo $M$}{
            Update weight: $\boldsymbol{w}_{t+1}(m) \gets \boldsymbol{w}_t(m)\exp\left(\alpha_t\, \mathbbm{1}\left(y^{(m)} \neq h_S^{(t)}(\rho^{(m)})\right)\right)$\;
        }
        Normalize weights: $\boldsymbol{w}_{t+1}(m) \gets \frac{\boldsymbol{w}_{t+1}(m)}{\sum_{m=1}^M \boldsymbol{w}_{t+1}(m)}$ for all $m$\;
    }
    Construct final classifier: $H_S(\cdot) \gets \arg\max_{k \in [K]} \sum_{t=1}^T \alpha_t\, \mathbbm{1}(h_S^{(t)}(\cdot) = k)$\;
    \Return{$H_S$}
\end{algorithm}

\resetAppendixCounters{C}

\section{The Reason for Using SGD in Classical Neural Networks}
\label{app:sgd_reason}

\begin{figure}[htpb]
    \centering
    \includegraphics[width=0.9\textwidth]{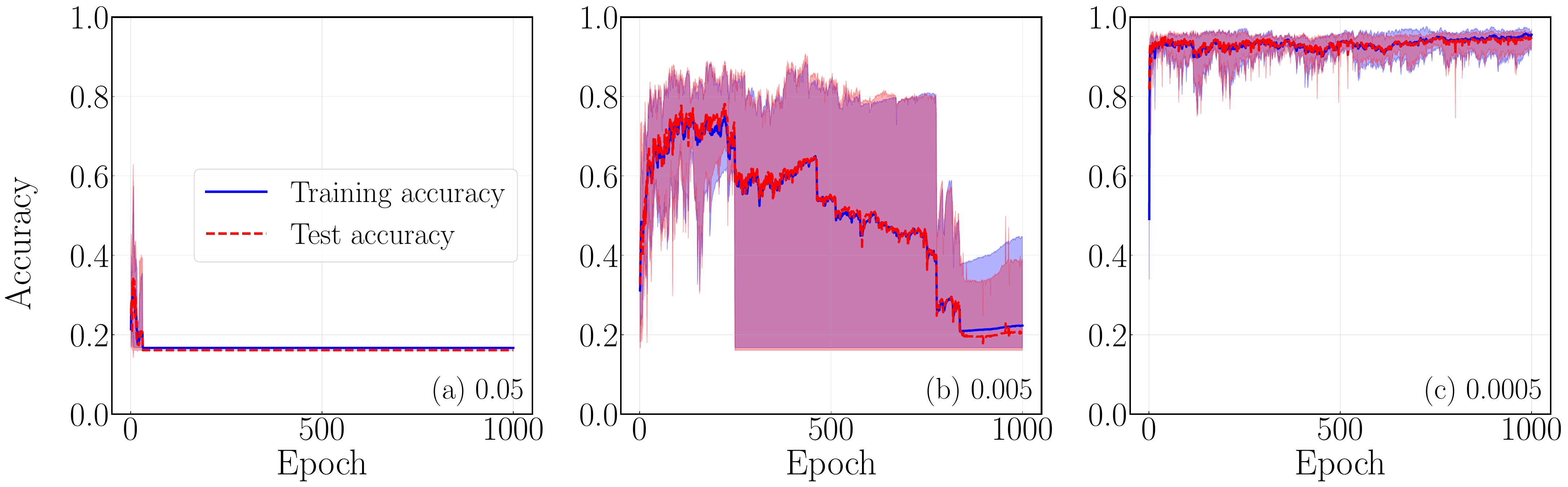}
    \caption{(a) Training and test accuracy of Adam on ViT with learning rate 0.05, (b) training and test accuracy with learning rate 0.005, and (c) training and test accuracy with learning rate 0.0005. The center line represents the average value over 5 runs and the shaded area indicates the minimum and maximum across those 5 runs.}
    \label{fig:adam_vit}
\end{figure}

In our experiments, we used the SGD optimizer to train ResNet and ViT. There are two main reasons for choosing SGD: on the one hand, previous research has shown that, compared to Adam, SGD tends to result in better generalization performance~\cite{zhou2020towards}; on the other hand, we found that in our experimental setting, the Adam optimizer could not achieve satisfactory training results for the ViT small model.

Specifically, we trained the ViT small model using the Adam optimizer with learning rates of 0.05, 0.005, and 0.0005, respectively. The maximum number of training epochs was set to 1000, and early stopping was applied if the train accuracy reached 100\%. All other parameter settings followed those in the main text. The training results are shown in Fig.~\ref{fig:adam_vit}. When the learning rate was set to 0.05, an abnormal phenomenon was observed: as the number of training epochs increased, the accuracies on both the train and test sets actually decreased. When using a learning rate of 0.005 (which is also the setting used for all Adam experiments in the main text), the accuracy rose at first but then declined, never surpassing 0.9. When the learning rate was further reduced to 0.0005, the accuracy could steadily exceed 0.9, but never reached 100\%, and overall performance remained inferior to SGD. These results further demonstrate that, under our experimental conditions, the SGD optimizer provides more stable and superior training outcomes than Adam.

\resetAppendixCounters{D}

\section{Details on Early Stopping Strategies}\label{sec:early_stopping_details}

\begin{figure}[htpb]
    \centering
    \includegraphics[width=0.8\textwidth]{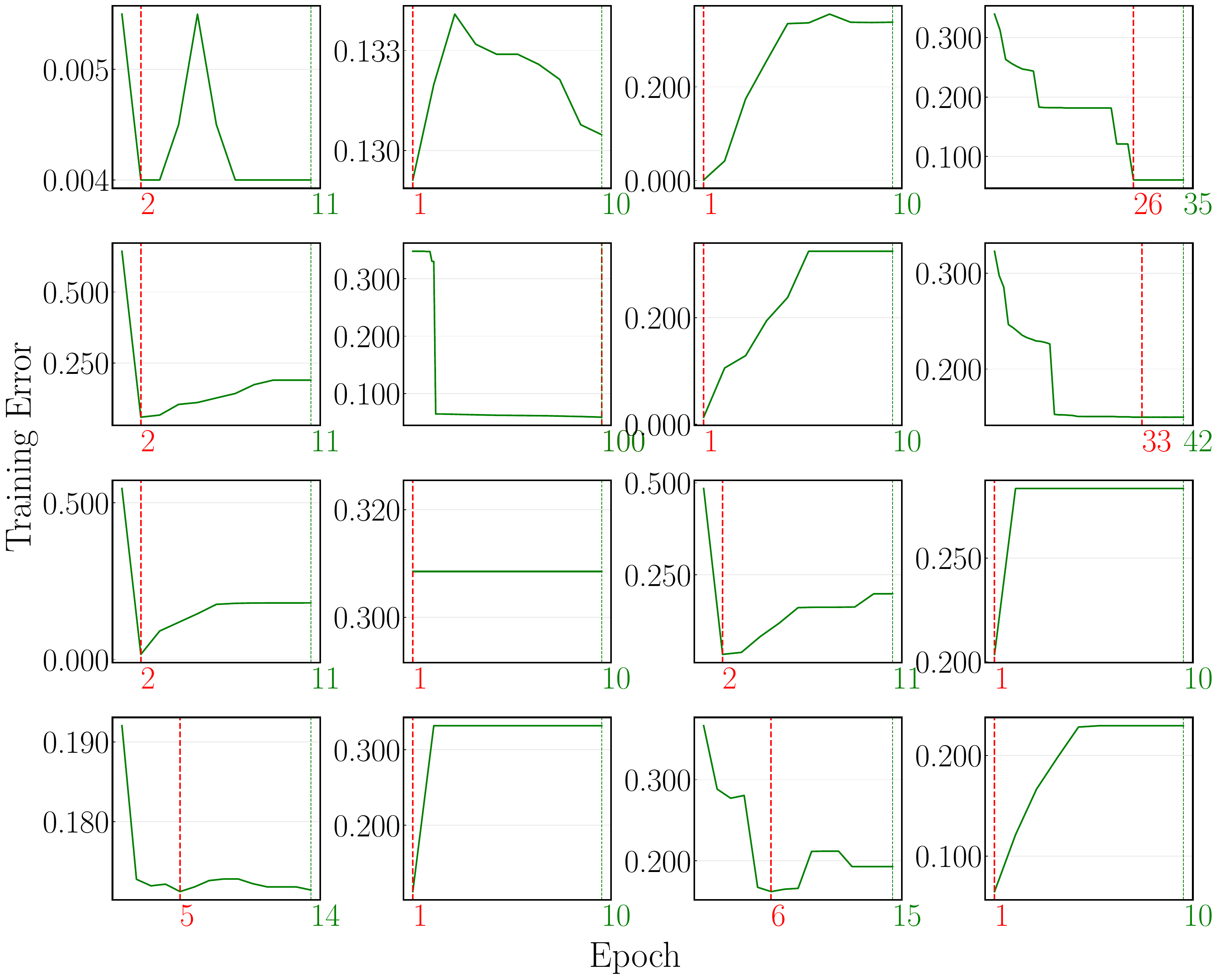}
    \caption{Train error of each base classifier when early stopping is applied. The red vertical line indicates the epoch at which the optimal train error is achieved and the corresponding parameters are selected. The green vertical line indicates the epoch when the early stopping strategy takes effect and training is finally halted. All subplots are arranged from 1 to 16, from left to right and top to bottom.}
    \label{afig:early}
\end{figure}

\begin{figure}[htpb]
    \centering
    \includegraphics[width=0.8\textwidth]{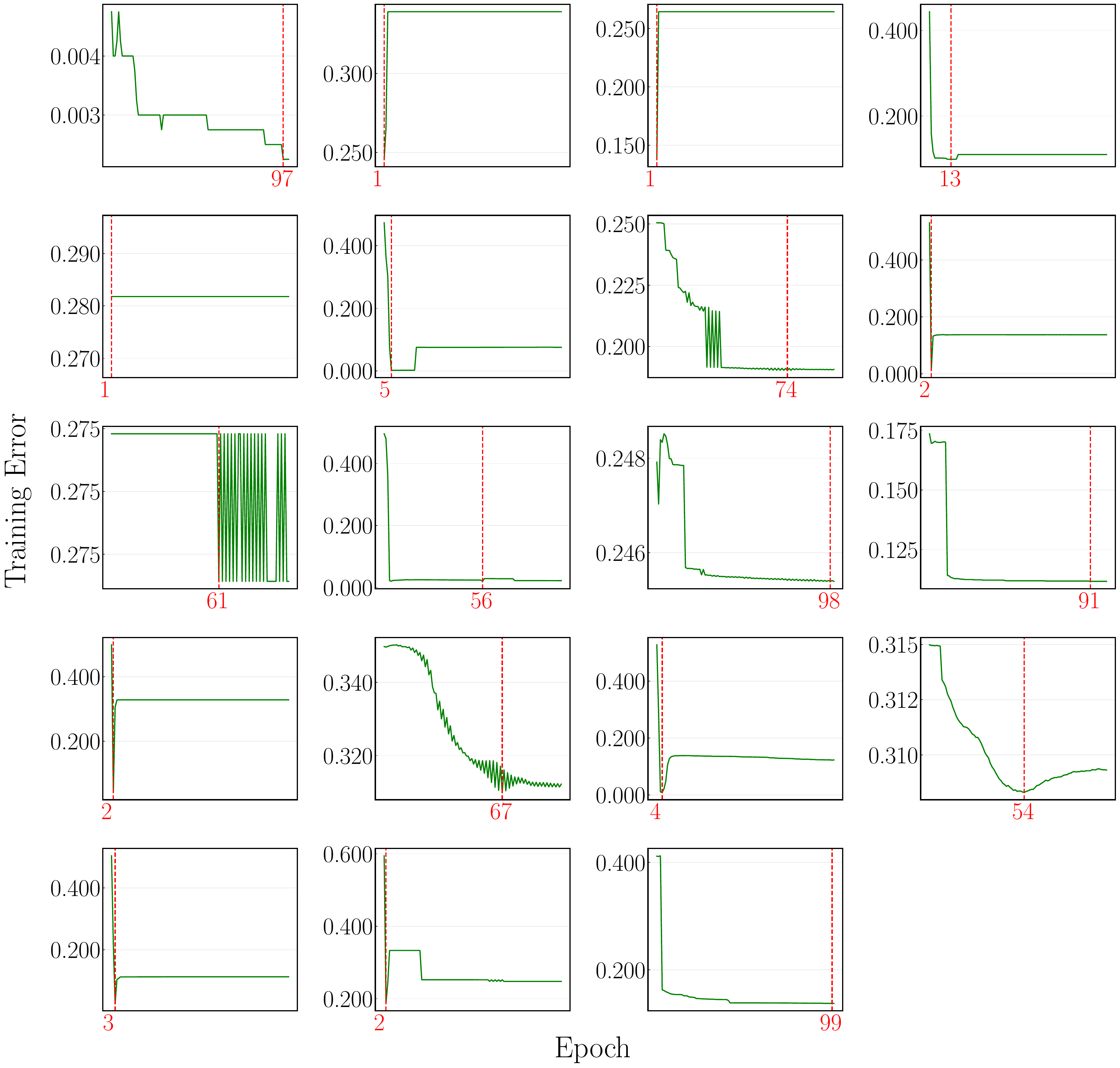}
    \caption{Train error of each base classifier when early stopping is not applied. The red vertical line indicates the epoch at which the optimal train error is achieved and the corresponding parameters are selected. All subplots are arranged from 1 to 19, from left to right and top to bottom.}
    \label{afig:no_early}
\end{figure}

In the main text, we have pointed out that the early stopping strategy can reduce training resources without sacrificing training effectiveness. Here, we provide a detailed demonstration, as shown in Figures~\ref{afig:early} and~\ref{afig:no_early}, of the performance of the $t$-th base classifier on the FashionMNIST (classes 1 and 9) training set with reweighted samples, under both cases of applying and not applying early stopping.

\end{document}